\documentclass[fleqn,twocolumn]{article}
\usepackage{geometry}
\makeatletter
\renewcommand{\fnum@figure}{Fig. \thefigure}
\makeatother

\usepackage[noblocks]{authblk}
\usepackage[dvipsnames]{xcolor}
\definecolor{mygreen}{RGB}{50, 180, 80}
\definecolor{myblue}{RGB}{50, 80, 180}
\usepackage[hidelinks]{hyperref}
\hypersetup{
	colorlinks   = true,
	urlcolor     = myblue,
	linkcolor    = myblue,
	citecolor   = mygreen
}
\usepackage[pdftex]{graphicx}
\graphicspath{{Figs/}}
\DeclareGraphicsExtensions{.pdf,.jpg,.png,.eps}
\usepackage{pifont}
\usepackage{upgreek}
\usepackage{cite}
\usepackage{amssymb}
\usepackage{amsmath}
\usepackage{multirow}
\usepackage{float}
\usepackage{amsfonts}
\usepackage[numbers,sort,compress]{natbib}

\begin{document}

\title{\textbf{An Analytical Strategy for Passive Harmonic Filter Placement in Transmission Systems with  Stochastic Aggregate Load Models Considering Resonant Conditions}}                      

\author[1]{Behnam Akbari\thanks{behnam.akbari@alum.sharif.ir}}
\author[1]{Farhad Pourtahmasbi\thanks{fpourtahmasbi@alum.sharif.edu}}
\author[1]{Hossein Mokhtari\thanks{mokhtari@sharif.edu}}

\affil[1]{Department of Electrical Engineering, Sharif
	University of Technology, Tehran, Iran}

\date{}

\maketitle

\begin{abstract}
{\small High percentage of voltage harmonics has been observed in transmission networks due to  harmonic currents penetrated from the load side amplified by resonant conditions. This requires the use of suitable harmonic filters in transmission networks. However, filter placement in a transmission system for harmonic mitigation is a planning procedure of high complexity. Since filter placement discounting the overall system response may deteriorate harmonic conditions at remote buses due to frequency characteristics alteration. The stochastic behavior of harmonic loads demands adequate models and calculations. The interlinked topology of a transmission system also necessitates an extensive system-wide study. This paper presents an analytical methodology for determining the appropriate number, location, and parameters of C-Type passive filters in a transmission system to improve the overall harmonic condition. The proposed strategy targets minimizing the expected value of a system-wide index given the stochastic behavior of loads, while limiting the 95\textsuperscript{th} percentiles of voltage distortions to the standard levels prescribed by IEEE Std. 519 and satisfying the fundamental power flow constraints. A restricted hierarchical direct search algorithm is utilized to find the near global solution whose subsets are also effective. The performance of the proposed approach is evaluated on the IEEE 118-bus test system and its validity is checked by Monte Carlo simulations and compared with harmonic modal analysis.}\end{abstract}


\section{Introduction}
Extensive use of nonlinear loads brings about considerable amounts of harmonic currents propagating into the utility grid. As a consequence, bus voltages become distorted and the power network is adversely affected. To avoid the adverse consequences of harmonic distortions, miscellaneous harmonic mitigation techniques have been introduced among which passive filter placement is an effective yet affordable strategy.

Since most nonlinear loads are connected at the distribution level, harmonic compensation at the same level is widely studied in the literature. Harmonic compensation schemes in distribution systems are focused on restricting distortions to the limits stipulated by standards and grid codes rather than totally eliminating the harmonic components. As a result, harmonic currents may spread into transmission systems and even be magnified due to impedance resonance characteristics. Real cases have been observed where the voltage harmonic distortion at high-voltage (HV) buses greatly exceeds the standard limit in practice. Therefore, constraining the harmonic distortion levels in distribution systems to standard limits, does not guarantee acceptable harmonic voltage levels in transmission systems. Especially when the nonlinear loads are scattered, filter placement at higher voltage levels becomes indispensable. The use of passive filter(s) in transmission networks can alleviate this problem by changing the flow of harmonic currents and harmonic impedance characteristics. This harmonic mitigation is feasible if the passive filters are properly allocated. Otherwise, the passive filter may reduce the harmonic level at its location but magnify at other buses in the power system. To the authors' best knowledge, limited research has been done so far to strategically address harmonic complications in transmission systems using passive filter placement.

The harmonic filter solution should ideally produce an improvement in the overall system harmonic condition. In order to quantify the overall system harmonic condition, a system-wide index is required. Such an index is introduced in \cite{Grady1992} to analyze the harmonic effects of installing an active power line conditioner in a test system. The same index is used in \cite{Chang2004, Chang2006, Belchior2015} for optimal placement of passive filters. Similar indices may be defined for harmonic studies, such as the one used in \cite{Chang2002} for passive filter planning.

References \cite{Grady1992,Chang2004,Chang2006,Belchior2015,Chang2002} have applied their optimization algorithms to distribution networks and have proposed a single-tuned filter topology for harmonic mitigation. However, for a transmission system of HV level with harmonic complications, the C-type filter solution is more practical, as the single-tuned filter is inefficient in mitigating voltage harmonics with frequencies of much higher than the filter tuned frequency \cite{Chou2000}. Additionally, a single-tuned filter may cause a greater magnification in the harmonics with frequencies of lower than the filter tuned frequency \cite{Chou2000}. C-type filters are also used in transmission systems by utilities to avoid undesirable resonances, while having negligible power losses of the fundamental frequency in comparison with other filter topologies \cite{Xu2016}.

Harmonic filter placement studies \cite{Chang2004,Chang2006,Cabral2017,Ortmeyer1996,Jannesar2018} typically assume a deterministic harmonic injection at system buses. Nevertheless, this assumption is not realistic especially at the transmission level where the harmonic injections are the aggregate equivalents of numerous nonlinear loads. Alternative methods have been introduced in the literature which do not require any knowledge about the system nonlinear loads. One popular method is the modal analysis which is expounded on in \cite{Xu2005}. Using modal analysis, one can identify the inherent harmonic resonances in a network and the buses that can potentially excite these resonances. The sensitivity of the resonance modes to network elements is further examined in \cite{Huang2007}. The studies \cite{Xu2005,Huang2007} solely take the system impedance characteristic into account and disregard the effect of the system nonlinear loads. In practice, the resonance modes is no menace unless excited. Reference \cite{Au2007b} has introduced a modal-based sensitivity index that includes the effects of both system impedances and harmonic injections. This index, whose calculation does not require comprehensive knowledge about the loads, is used for passive filter placement. The limitation about this approach is that it is only applicable to radial distribution systems.

Harmonic currents of aggregate harmonic loads behave stochastically due to various factors, e.g. random variation of nonlinear load composition and random operating conditions of individual nonlinear loads \cite{Au2007}. To account for this stochasticity, a comprehensive current injection model is proposed in \cite{Au2007} and used in \cite{Au2007a} for calculating the voltage harmonic in a distribution network. A similar stochastic model is introduced in \cite{Silva2015} and applied to a distribution feeder using Monte Carlo simulations (MCS) to predict its harmonic distortion levels. Another approach is presented in \cite{Hong2013} for considering the stochastic behavior of harmonic loads in planning single-tuned filters. The application of analysis procedures proposed in \cite{Au2007a,Silva2015,Hong2013} for transmission harmonic studies has not been investigated.

In this paper, a framework is presented for harmonic studies in a network whose loads are stochastically modeled. For a broad overview, a system-wide index is introduced based on previous studies which emphasizes the buses with higher distortion levels. Analytical formulations are derived for estimating the distribution of this index in the presence of probabilistic loads. The formulations are put into matrix forms to enhance execution speed in MATLAB software. Based on the presented framework, a strategy is proposed for optimal passive filter placement in transmission systems. The proposed strategy is used to allocate C-type filters in IEEE 118-bus test system in order to maximize the expected value of the harmonic index and to meet standard requirements. The accuracy of the results is examined using MCS. Harmonic modal analysis is also used to analyze how the proposed filter placement strategy affects the system impedance characteristics.

\section{Load Model}
\begin{figure}
	\centering
	\includegraphics[width=3.2in]{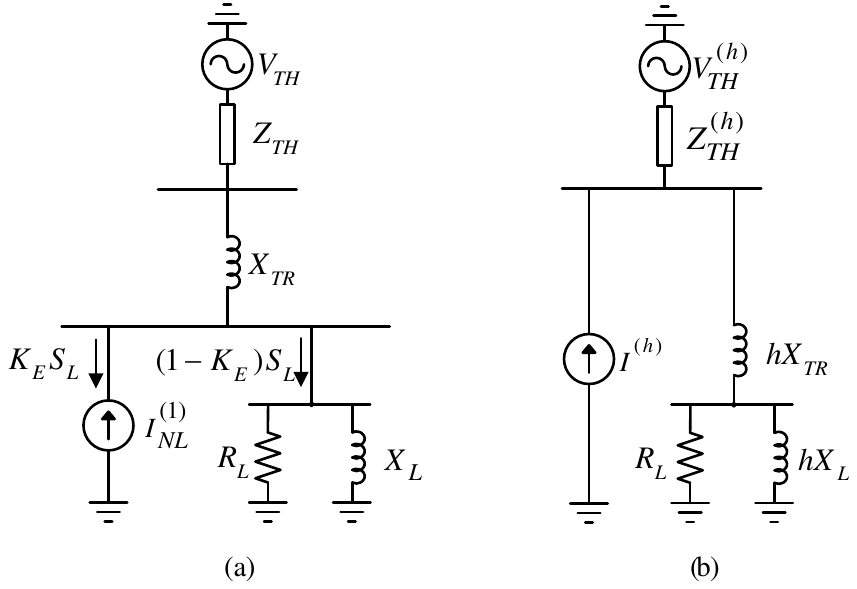}
	\caption{Aggregate harmonic load model connected to system thevenin equivalent represented at (a) the fundamental harmonic and (b) an arbitrary harmonic.}
	\label{fig.Load}
\end{figure}

An aggregate harmonic load connected to the system thevenin impedance ($Z_{\textit{TH}}$) can be modeled as
\hyperref[fig.Load]{Fig.~\ref*{fig.Load}~(a)} \cite{Au2007a}, where $X_{\textit{TR}}$ represents the transformer impedance, the current source models the current injection of nonlinear loads, the parallel RL impedance models the behavior of linear loads, and $K_E$ is the ratio of the apparent power of nonlinear loads to that of the total load. The linear load and transformer impedances are absorbed in the system impedance matrix for harmonic studies, and the harmonic injection $I_{\textit{NL}}^{(h)}$ can be transferred to the system-side as depicted in \hyperref[fig.Load]{Fig.~\ref*{fig.Load}~(b)} according to:
{\setlength{\mathindent}{0cm}\begin{equation}\label{eq.ILN}
	\begin{split}&  \frac{Z_L^{(h)}}{Z_L^{(h)}+jX_{\textit{TR}}^{(h)}+Z_{\textit{TH}}^{(h)}} I_{\textit{NL}}^{(h)} = \frac{Z_L^{(h)}+jX_{\textit{TR}}^{(h)}}{Z_L^{(h)}+jX_{\textit{TR}}^{(h)}+Z_{\textit{TH}}^{(h)}} I^{(h)}\\
	& \Rightarrow I^{(h)}=\frac{Z_L^{(h)}}{Z_L^{(h)}+jX_{\textit{TR}}^{(h)}} I_{\textit{NL}}^{(h)}\end{split}\end{equation}}

\noindent where $Z_L$ represents the impedance of the linear portion of the load.

The harmonic injection at an HV bus is the vector sum of the currents of numerous nonlinear loads. As a result, this injection amplitude is not usually deterministically known, unless accurately measured. There is even more uncertainty about harmonic injection phase, since most industrial measurement devices do not report the phase angles between the harmonic currents and a global reference. Accordingly, a stochastic representation of the harmonic injections in the following form can be beneficial for harmonic studies:
\nopagebreak
{\setlength{\mathindent}{0cm}\begin{equation}\label{eq.load}
	I_j^{(h)}=\alpha_j^{(h)}|I_j^{(1)}|e^{i ( h  \angle I_{j}^{(1)} + \phi_j^{(h)})}
	\end{equation}}

\noindent where $\alpha_j^{(h)}$ is the ratio of the amplitude of the $h$\textsuperscript{th} harmonic current to the fundamental current of the nonlinear load at bus $j$ and $\phi_j^{(h)}$ is the relative angle of the harmonic current. In this stochastic representation, $\alpha_j^{(h)}$ and $\phi_j^{(h)}$ are random variables.

\section{C-type Filter}
\label{S3}
\begin{figure}
	\centering
	\includegraphics[width=1in]{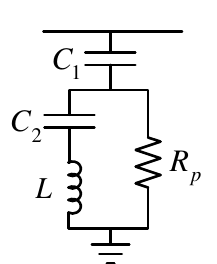}
	\caption{Single-line diagram of a C-type harmonic filter.}
	\label{fig.C-type}
\end{figure}

A C-type filter circuit model is depicted in \hyperref[fig.C-type]{Fig.~\ref*{fig.C-type}}. This representation has four passive elements. In order to minimize the power losses in the parallel resistance ($R_p$), the elements $C_2$ and $L$ should resonate at the fundamental frequency \cite{Bodger1990}. In other words, the reactance magnitudes of $C_2$ and $L$ are equal at the fundamental frequency. Given this equality, the filter is fully known if the following parameters are determined.

\subsection{Fundamental Impedance ($Z_b$)}
The fundamental impedance for a C-type filter is determined by the reactive capacity of the filter at the fundamental frequency.
\nopagebreak
{\setlength{\mathindent}{0cm}\begin{equation}\label{eq.Zb}
	Z_b = \frac{V^2}{Q}
	\end{equation}}

\subsection{Tuning Harmonic Order ($h_t$)}
Provided that the parallel resistance is infinite, the tuning frequency of the filter, represents the frequency at which the filter's impedance is minimum.
\nopagebreak
{\setlength{\mathindent}{0cm}\begin{equation}\label{eq.ht}
	h_t =  \frac{1}{\omega_0}\sqrt{\frac{C_1+C_2}{C_1 C_2 L}}
	\end{equation}}

\subsection{Quality Factor ($q$)}
The quality factor of the C-type filter is a measure of the damping resistance. A C-type filter with a high $q$ has an identical frequency response to a single-tuned filter, i.e.:
\nopagebreak
{\setlength{\mathindent}{0cm}\begin{equation}\label{eq.q}
	q = h_t \frac{R_p}{Z_b}\textnormal{.}
	\end{equation}}

The filter impedance at an arbitrary harmonic can be calculated as:
\nopagebreak
{\setlength{\mathindent}{0cm}\begin{equation}\label{eq.Zf}
	Z_f^{(h)} = Z_b [\frac{q(h^2-1)[h_t(h^2-1)+jqh(h_t^2-1)]}{h_t^2(h^2-1)^2+qh^2(h_t^2-1)^2}-\frac{j}{h}]\textnormal{.}
	\end{equation}}

$h_t$ is usually set below the lowest expected resonant frequency of the system \cite{Horton2012}. In many of the C-type filters used in HVDC projects, $h_t$ is considered to be $3$ \cite{YaoXiao2004}. Additionally,  $h_t$ should be sufficiently larger than $1$ in order to avoid unintended resonances in the impedance characteristics. Also, the quality factor is practically selected in the range of $ 1 \leq q \leq 2.3 $ \cite{YaoXiao2004}.

\begin{figure}
	\centering
	\includegraphics[width=3.2in]{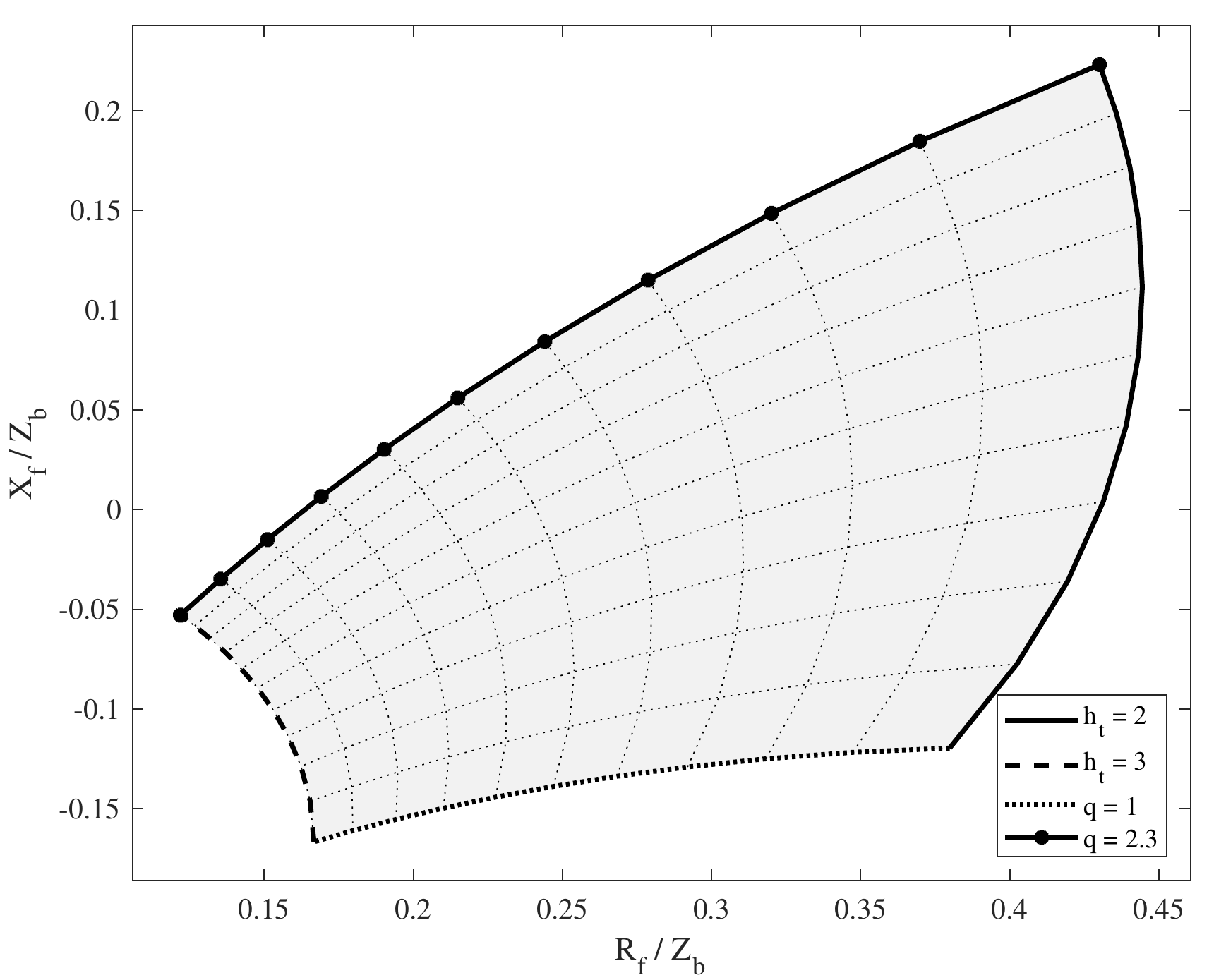}
	\caption{C-type filter impedance region at the 3\textsuperscript{rd} harmonic.}
	\label{fig.locus_h3}
\end{figure}

\begin{figure}
	\centering
	\includegraphics[width=3.2in]{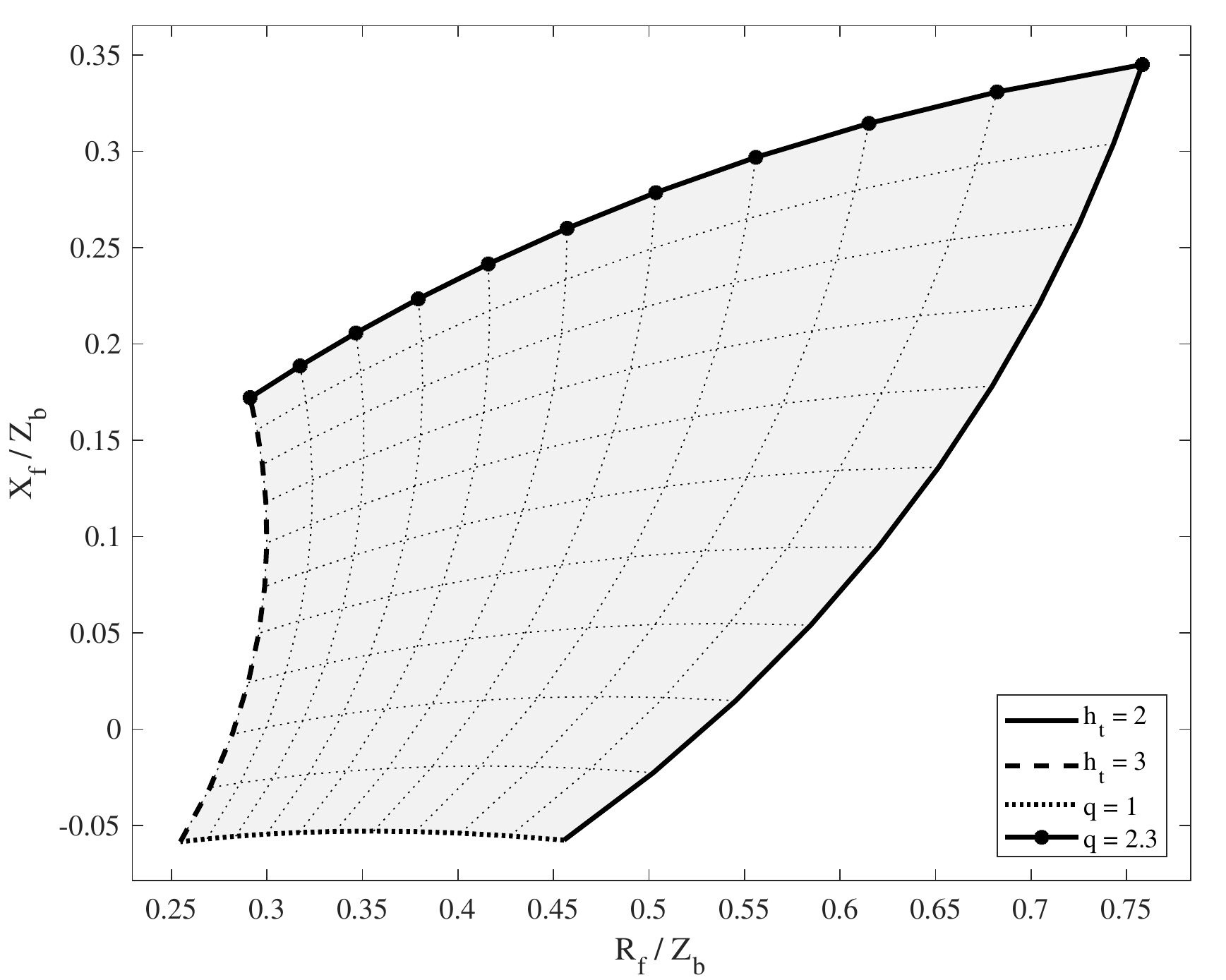}
	\caption{C-type filter impedance region at the 5\textsuperscript{th} harmonic.}
	\label{fig.locus_h5}
\end{figure}

The feasible regions for the filter impedance considering the stated limits at harmonics 3 and 5 are depicted in \hyperref[fig.locus_h3]{Fig.~\ref*{fig.locus_h3}} and \hyperref[fig.locus_h5]{Fig.~\ref*{fig.locus_h5}}. The filter impedance at these harmonics decreases as $h_t$ increases. A filter with a smaller impedance magnitude at a given harmonic order is usually more capable of suppressing the voltage harmonics of the same order. As a result, $h_t$ is selected to be $3$. It is concluded from these figures that at the 3\textsuperscript{rd} harmonic, the filter with the largest $q$, and at the 5\textsuperscript{th} harmonic, the filter with the smallest $q$, have the lowest impedances. For higher order harmonics, the trend is the same as the 5\textsuperscript{th} harmonic. Thus there is a compromise for selecting the optimum $q$ according to the harmonic components of the grid.

\section{Methodology}
\label{S4}

\subsection{Analytical Framework for Harmonic Studies}
\label{S2.A}

The harmonic power flow is governed by the following equation, which yields the harmonic voltage at bus $k$:
{\setlength{\mathindent}{0cm}\begin{equation}\label{eq.VIHD0}
V_k^{(h)} =  \left|\sum_{j}^{N}{Z_{ij}^{(h)}I_{j}^{(h)}}\right|
\end{equation}}

\noindent where $Z_{ij}^{(h)}$ is the transfer impedance from bus $i$ to bus $j$ at harmonic $h$, $I_j^{(h)}$ is the $h$\textsuperscript{th} harmonic injection at bus $j$, and $N$ is the number of system buses.

In order to derive the harmonic power flow equations based on the aforementioned load model, the following variables are defined:
{\setlength{\mathindent}{0cm}\begin{equation}\label{eq.U}
U_{ij}^{(h)}=\frac{\left| Z_{ij}^{(h)}I_{j}^{(1)} \right|}{V_{i}^{rated}},\text{ }\mathbf{U}_{N\times N}^{(h)}=\left[ U_{ij}^{(h)} \right]\textnormal{,}
\end{equation}}

\noindent and 
{\setlength{\mathindent}{0cm}\begin{equation}\label{eq.theta}
\theta _{ij}^{(h)}=\measuredangle Z_{ij}^{(h)}+h\measuredangle I_{j}^{(1)},\text{ }\boldsymbol{\uptheta }_{N\times N}^{(h)}=\left[ {{e}^{i\theta _{ij}^{(h)}}} \right]\textnormal{.}
\end{equation}}

According to IEEE Std. 519 \cite{IEEE519}, the voltage harmonic distortions should be expressed in percentage of the rated power-frequency voltage. Therefore, the voltage individual harmonic distortion (IHD) and total harmonic distortion (THD) at the buses can be calculated as:
{\setlength{\mathindent}{0cm}\begin{equation}\label{eq.VIHD}
\begin{split}
V{{_\textit{IHD}^{(h)}}_{,k}}=\left| \sum_{j}^{N}{U_{kj}^{(h)}\alpha _{j}^{(h)}}{{e}^{i\left( \theta _{kj}^{(h)}+\phi _{j}^{(h)} \right)}} \right|, \\
{\mathbf{V}}_{{\textit{IHD}}_{N\times 1}}^{(h)}={{\left[ \begin{matrix}
		V_{\textit{IHD},1}^{(h)} & \cdots  & V_{\textit{IHD},N}^{(h)}  \\
		\end{matrix} \right]}^{T}}\textnormal{,}
\end{split}
\end{equation}}

\noindent and
{\setlength{\mathindent}{0cm}\begin{equation}\label{eq.VTHD}
\mathbf{V}_{\textit{THD}}^{\circ 2}=\sum\limits_{h}{(\mathbf{V}_{\textit{IHD}}^{(h)})^{\circ 2}}
\end{equation}}

\noindent where $\circ$ is the Hadamard operator.

All individual and total harmonic distortions may be expressed in a single matrix as:
{\setlength{\mathindent}{0cm}\begin{equation}\label{eq.VHD}
\mathbf{V}_{\textit{HD}}=\left[ \begin{matrix}
\mathbf{V}_{\textit{IHD}}^{(h_1)} & \mathbf{V}_{\textit{IHD}}^{(h_2)} & \cdots  & \mathbf{V}_{\textit{THD}}  \\
\end{matrix} \right]\textnormal{.}
\end{equation}}

Hereafter, the superscript ${(h)}$ is dropped for conciseness. The average of vector $\mathbf{V}_{\textit{THD}}^{\circ 2}$ over the system buses has been defined in \cite{Grady1992,Chang2004,Chang2006,Belchior2015} as a system-wide harmonic index. The same index is used in this paper, except the index is a random variable here, as the harmonic loads are stochastically modeled. The individual and total harmonic system-wide indices are mathematically expressed as:
{\setlength{\mathindent}{0cm}\begin{equation}\label{eq.SIHD}
{{S}_{\textit{IHD}}}=\frac{1}{N}{{\mathbf{e}}^{T}}\mathbf{V}_{\textit{IHD}}^{\circ 2}\textnormal{.}
\end{equation}}

\noindent and
{\setlength{\mathindent}{0cm}\begin{equation}\label{eq.STHD}
{{S}_{\textit{THD}}}=\sum\limits_{h}{{{S}_{\textit{IHD}}}}=\frac{1}{N}{{\mathbf{e}}^{T}}\mathbf{V}_{\textit{THD}}^{\circ 2}
\end{equation}}
\noindent where $\mathbf{e}$ is an $N \times 1$ all-ones vector. 

The means of \hyperref[eq.SIHD]{(\ref*{eq.SIHD})} and \hyperref[eq.STHD]{(\ref*{eq.STHD})} can properly represent the harmonic status of the system emphasizing the critical system buses. For reference, the means and variances of \hyperref[eq.VIHD]{(\ref*{eq.VIHD})} through \hyperref[eq.STHD]{(\ref*{eq.STHD})}, which include random variables $\alpha_j$ and $\phi_j$, are found. In this regard, the following matrices and vectors are defined.

{\setlength{\mathindent}{0cm}\begin{equation}\label{eq.alpha}
\boldsymbol{\upalpha }_{N\times 1}={{\left[ \begin{matrix}
		\alpha _{1} & \cdots  & \alpha _{N}  \\
		\end{matrix} \right]}^{T}}
\end{equation}}
{\setlength{\mathindent}{0cm}\begin{equation}\label{eq.phi}
\boldsymbol{\upphi }_{N\times 1}={{\left[ \begin{matrix}
		{{e}^{i\phi _{1}}} & \cdots  & {{e}^{i\phi _{N}}}  \\
		\end{matrix} \right]}^{T}}
\end{equation}}
{\setlength{\mathindent}{0cm}\begin{equation}\label{eq.UcAc}
{{\mathbf{U}}_{c}}=\mathbf{U}\circ \boldsymbol{\uptheta },\text{ }{{\boldsymbol{\upalpha }}_{c}}=\boldsymbol{\upalpha }\circ \boldsymbol{\upphi }
\end{equation}}
{\setlength{\mathindent}{0cm}\begin{equation}\label{eq.mu2}
{{{\mu }}_{2,j}'}=\operatorname{E}\left[ \alpha _{j}^{2} \right],\text{ }{{\boldsymbol{{\upmu }}}_{{{2}_{N\times 1}}}'}={{\left[ \begin{matrix}
		{{{{\mu }}}_{2,1}'} & \cdots  & {{{{\mu }}}_{2,N}'}  \\
		\end{matrix} \right]}^{T}}
\end{equation}}
{\setlength{\mathindent}{0cm}\begin{equation}\label{eq.mu4}
{{{\mu }}_{4,j}'}=\operatorname{E}\left[ \alpha _{j}^{4} \right],\text{ }{{\boldsymbol{{\upmu }}}_{{4}_{N\times 1}}'}={{\left[ \begin{matrix}
		{{{{\mu }}}_{4,1}'} & \cdots  & {{{{\mu }}}_{4,N}'}  \\
		\end{matrix} \right]}^{T}}
\end{equation}}
{\setlength{\mathindent}{0cm}\begin{equation}\label{eq.M}
{{\mathbf{M}}_{N\times N}}=\operatorname{diag}\left( {\boldsymbol{\upmu}_{2}'} \right)=\left[ \begin{matrix}
{{{{\mu }}}_{2,1}'} & 0 & \cdots  & 0  \\
0 & {{{{\mu }}}_{2,2}'} & \cdots  & 0  \\
\vdots  & \vdots  & \ddots  & \vdots   \\
0 & 0 & \cdots  & {{{{\mu }}}_{2,N}'}  \\
\end{matrix} \right]
\end{equation}}
{\setlength{\mathindent}{0cm}\begin{equation}\label{eq.Mp}
{{\mathbf{{M}}}_{N\times N}'}=\left[ {{m}_{ij}} \right],\text{ }{{m}_{ij}}=\left\{ \begin{matrix}
{{{{\mu }}}_{4,j}'} & i=j  \\
{{{{\mu }}}_{2,i}'}{{{{\mu }}}_{2,j}'} & i\ne j  \\
\end{matrix} \right.
\end{equation}}
{\setlength{\mathindent}{0cm}\begin{equation}\label{eq.Up}
{{\mathbf{{U}}}_{c}'}={{\mathbf{U}}_{c}}\mathbf{{M}'}
\end{equation}}

If  $ \phi_j \sim \textrm{unif}[0,2\pi] $, the squared value of \hyperref[eq.VIHD]{(\ref*{eq.VIHD})} can be written as:
{\setlength{\mathindent}{0cm}\begin{equation}\label{eq.VIHD2}
V_{\textit{IHD},k}^{2}=\sum_{i}^{N}{\sum_{j}^{N}{{{U}_{ki}}}}{{U}_{kj}}{{\alpha }_{i}}{{\alpha }_{j}}\cos \left( {{\theta }_{ki}}-{{\theta }_{kj}}+{{\phi }_{i}}-{{\phi }_{j}} \right)\textnormal{,}
\end{equation}}
{\setlength{\mathindent}{0cm}\begin{equation}\label{eq.VIHD3}
\mathbf{V}_{\textit{IHD}}^{\circ 2}=\operatorname{Re}\left\{ \left( {{\mathbf{U}}_{c}}{{\boldsymbol{\upalpha }}_{c}} \right)\circ \operatorname{conj}\left( {{\mathbf{U}}_{c}}{{\boldsymbol{\upalpha }}_{c}} \right) \right\}\textnormal{.}
\end{equation}}

The expected value and variance of \hyperref[eq.VIHD3]{(\ref*{eq.VIHD3})} can be calculated as:
{\setlength{\mathindent}{0cm}\begin{equation}\label{eq.EVIHD}
\operatorname{E}[\mathbf{V}_{\textit{IHD}}^{\circ 2}]={{\mathbf{U}}^{\circ 2}}{{\boldsymbol{{\upmu}}}_{2}'}\textnormal{,}
\end{equation}}

\noindent and
{\setlength{\mathindent}{0cm}\begin{equation}\label{eq.VarVIHD}
\begin{split}
& \operatorname{Var}[\mathbf{V}_{\textit{IHD}}^{\circ 2}]=2\operatorname{diag}{{\left( \mathbf{UM}{{\mathbf{U}}^{T}} \right)}^{\circ 2}}\\
& +{{\mathbf{U}}^{\circ 4}}\left( {{{\boldsymbol{{\upmu }'}}}_{4}}-2\boldsymbol{{\upmu }'}_{2}^{\circ 2} \right)-{{\left( {{\mathbf{U}}^{\circ 2}}{{{\boldsymbol{{\upmu }'}}}_{2}} \right)}^{\circ 2}}\textnormal{.}
\end{split}
\end{equation}}

The expected values and variances of the squared total harmonic distortions are simply obtained as:
{\setlength{\mathindent}{0cm}\begin{equation}\label{eq.EVTHD}
\operatorname{E}[\mathbf{V}_{\textit{THD}}^{\circ 2}]=\sum\limits_{h}{\operatorname{E}[\mathbf{V}_{\textit{IHD}}^{\circ 2}]}\textnormal{,}
\end{equation}}

\noindent and
{\setlength{\mathindent}{0cm}\begin{equation}\label{eq.VarVTHD}
\operatorname{Var}[\mathbf{V}_{\textit{THD}}^{\circ 2}]=\sum\limits_{h}{\operatorname{Var}[\mathbf{V}_{\textit{IHD}}^{\circ 2}]}\textnormal{.}
\end{equation}}

The expected value and variance of \hyperref[eq.SIHD]{(\ref*{eq.SIHD})} are calculated as:
{\setlength{\mathindent}{0cm}\begin{equation}\label{eq.ESIHD}
\operatorname{E}[{{S}_{\textit{IHD}}}]=\frac{1}{N}{{\mathbf{e}}^{T}}{{\mathbf{U}}^{\circ 2}}{{\boldsymbol{{\upmu }'}}_{2}}\textnormal{,}
\end{equation}}
and
{\setlength{\mathindent}{0cm}\begin{equation}\label{eq.VarSIHD}
\begin{split}
& \operatorname{Var}\left[ {{S}_{\textit{IHD}}} \right]=\frac{1}{{{N}^{2}}} [ {{\mathbf{e}}^{T}}\left( \left( {{\mathbf{U}}^{\circ {{2}^{T}}}}\mathbf{e}\text{ }{{\mathbf{e}}^{T}}{{\mathbf{U}}^{\circ 2}} \right)\circ \mathbf{{M}'} \right)\mathbf{e}\\
& + \operatorname{Re}[ \operatorname{tr}\left( {{{\mathbf{{U}'}}}_{c}}^{*}{{{\mathbf{{U}'}}}_{c}}\mathbf{U}_{c}^{*}{{\mathbf{U}}_{c}}-\left( {{{\mathbf{{U}'}}}_{c}}^{*}{{{\mathbf{{U}'}}}_{c}} \right)\circ \left( \mathbf{U}_{c}^{*}{{\mathbf{U}}_{c}} \right) \right) ]\\
& -{{\left( {{\mathbf{e}}^{T}}{{\mathbf{U}}^{\circ 2}}{{{\boldsymbol{{\upmu}}}}_{2}'} \right)}^{2}} ]\textnormal{.}
\end{split}
\end{equation}}

Finally, the expected value and variance of \hyperref[eq.STHD]{(\ref*{eq.STHD})} are obtained as:
{\setlength{\mathindent}{0cm}\begin{equation}\label{eq.ESTHD}
\operatorname{E}\left[ {{S}_{\textit{THD}}} \right]=\sum\limits_{h}{\operatorname{E}\left[ {{S}_{\textit{IHD}}} \right]}\textnormal{,}
\end{equation}}

\noindent and
{\setlength{\mathindent}{0cm}\begin{equation}\label{eq.VarSTHD}
\operatorname{Var}\left[ {{S}_{\textit{THD}}} \right]=\sum\limits_{h}{\operatorname{Var}\left[ {{S}_{\textit{IHD}}} \right]}\textnormal{.}
\end{equation}}

\subsection{Harmonic Voltage Distribution Estimation}
\label{S2.B}
In order to evaluate the harmonic voltages, which are random variables, their distributions must be obtained. As the accurate calculation is not straightforward, one should estimate the distribution types and parameters.

Since $V^2_{\textit{HD}}$ is a skewed positive random variable, gamma or log-normal distributions can estimate its probability density \cite{Kocherlakota1995}. These distributions are often interchangeably used \cite{Wiens1999}.

If $\ln{(X)}$ has a normal distribution with a mean of $\mu$ and standard deviation of $\sigma$, $X$ has a log-normal distribution with parameters $\mu$ and $\sigma$ and its density function is:
\nopagebreak
{\setlength{\mathindent}{0cm}\begin{equation}\label{eq.fLN}
f_{\textit{LN}} (x;\mu,\sigma)=  \frac{1}{x\sigma\sqrt{2} \pi} e^{-\frac{\ln^2{(x-\mu)} }{2\sigma^2}} \quad x,\sigma > 0, \mu \in \mathbb{R}\textnormal{.}
\end{equation}}

Gamma probability density function (PDF) is denoted by:
\nopagebreak
{\setlength{\mathindent}{0cm}\begin{equation}\label{eq.fG}
f_{\textit{GA}}(x;\alpha,\theta)= \frac{1}{\Gamma(\alpha) \theta^\alpha} x^{\alpha-1} e^{\frac{-x}{\theta}} \quad x,\alpha,\theta>0
\end{equation}}

\noindent where $\alpha$ and $\theta$ are the shape and scale parameters respectively.

Some studies are conducted to discriminate between these distributions \cite{Wiens1999,Kundu2005}. Comparing the likelihoods is one approach for determining the preferred PDF \cite{Kundu2005}, which needs MCS. Considering that the data set $x=\{X_1...X_n\}$ is coming from a distribution with the density function $f(x;\textit{params})$, the likelihood function is defined as:
\nopagebreak
{\setlength{\mathindent}{0cm}\begin{equation}\label{eq.Likelihood}
L(\textit{params};x)=\prod\limits_{i=1}^{n}{f(X_i;\textit{params})}\textnormal{.}
\end{equation}}

For convenience, the natural logarithm of the likelihood, called log-likelihood, is often reported instead. In order to find the maximum likelihood fit (MLF) to a data set, the density function parameters are determined such that $L$ is maximized. However, in case the data set is difficult to obtain, its mean ($m$) and variance ($v$) may be used to estimate the distribution parameters. The gamma and log-normal distribution parameters of the estimated fit (EF) are computed according to:
{\setlength{\mathindent}{0cm}\begin{equation}\label{eq.gamma}
\hat{\theta} = \frac{v}{m}, \quad \hat{\alpha} = \frac{m^2}{v}\textnormal{,}
\end{equation}}

\noindent and
{\setlength{\mathindent}{0cm}\begin{equation}\label{eq.ln}
\hat{\mu} = \ln\left(\frac{m^2}{\sqrt{v+ m^2}}\right), \quad \hat{\sigma} = \sqrt{\ln(\frac{v}{m^2}+1)}\textnormal{.}
\end{equation}}

The mean and variance of $V^2_{\textit{HD}}$ are calculated according to \hyperref[eq.EVIHD]{(\ref*{eq.EVIHD})} through \hyperref[eq.VarVTHD]{(\ref*{eq.VarVTHD})}. Without using MCS distinguishing between these distributions is not possible. Therefore, the 95\textsuperscript{th} percentiles are found using both distributions and the maximum value is considered for conservativeness.

As $V_{\textit{HD}}$ is a positive random variable, its 95\textsuperscript{th} percentile is the square root of the  95\textsuperscript{th} percentile of $V^2_{\textit{HD}}$. Thus, the matrix $\mathbf{\tilde{V}}_{\textit{HD},95\%}$, which is the estimation for 95\textsuperscript{th} percentile of $\mathbf{V}_{\textit{HD}}$, can be analytically calculated.

\subsection{Filter Placement Strategy}
\label{S2.D}
\begin{figure}
	\centering
	\includegraphics[width=3.2in]{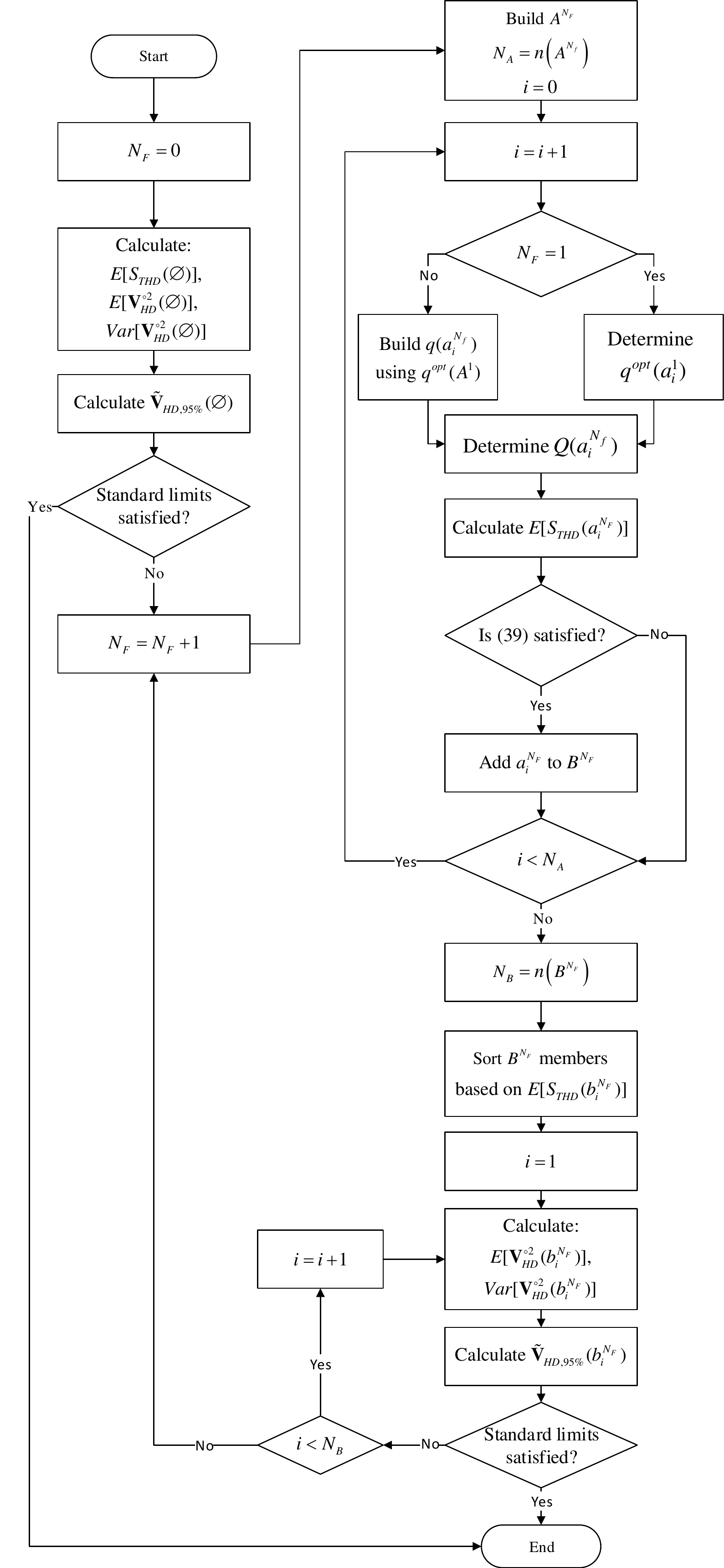}
	\caption{Filter placement strategy.}
	\label{fig.Flowchart}
\end{figure}

A hierarchical strategy is proposed for optimal filter placement with the steps described in the following and illustrated in \hyperref[fig.Flowchart]{Fig.~\ref*{fig.Flowchart}}. In this strategy, when a filter is located at a bus with an already available capacitor, the filter replaces the capacitor. For the reasons previously stated, the allocation of C-type filters is solely investigated in this paper. Though the strategy may be extended to include other filter topologies.

\subsubsection{Step 1}
The network is analyzed in the base case to calculate the expected value of the initial system-wide index ($\operatorname{E}[{{S}_{\textit{THD}}}(\varnothing)$). The expected value and variance for each element of the $\mathbf{V}_{\textit{HD}}^{\circ 2}$ matrix is found in order to estimate the 95\textsuperscript{th} percentile harmonic distortions with the approach explained in subsection~\ref{S2.B}. The matrix $\mathbf{\tilde{V}}_{\textit{HD},95\%}(\varnothing )$ is then formed.
As prescribed by IEEE Std. 519 \cite{IEEE519}, the 95\textsuperscript{th} percentile of voltage distortions at buses should be limited to the values expressed in \hyperref[table.VHD_limits]{Table~\ref*{table.VHD_limits}}. If there is no standard limit violation, the algorithm is terminated. Otherwise, the number of required filters is incremented to $N_F=1$ and the algorithm proceeds to \textbf{Step 2}.

\begin{table}
	\renewcommand{\arraystretch}{1.3}
	\caption{Voltage distortion limits \cite{IEEE519}}
	\label{table.VHD_limits}
	\centering
	\begin{tabular}{|c|c|c|}
		\hline
		Rated Voltage & $V_{\textit{IHD}}$ & $V_{\textit{THD}}$ \\
		\hline
		V $\leq$ 1.0 kV & 5.0 & 8.0 \\
		\hline
		1.0 kV $<$ V $\leq$ 69 kV & 3.0 & 5.0 \\
		\hline
		69 kV $<$ V $\leq$ 161 kV & 1.5 & 2.5 \\
		\hline
		161 kV $<$ V & 1.0 & 1.5 \\
		\hline
	\end{tabular}
\end{table}

\subsubsection{Step 2}
The set containing all possibly desirable $N_F$-tuple filter combinations ($A^{N_F}$) is built according to \hyperref[eq.A1]{(\ref*{eq.A1})} if $N_F = 1$ and \hyperref[eq.A]{(\ref*{eq.A})} if $N_F \geq 2$.
{\setlength{\mathindent}{0cm}\begin{equation}\label{eq.A1}
A^{1}=\left\{a^{1}\mid a^1 \text{ is a candidate single-filter scenario}\right\}
\end{equation}}
{\setlength{\mathindent}{0cm}\begin{equation}\label{eq.A}
A^{N_F}=\left\{a^{N_F}\mid \forall a^{N_F-1} \subset a^{N_F} : a^{N_F-1} \in B^{N_F-1}\right\}
\end{equation}}

\noindent where $B^{N_F-1}$ is the set containing all desirable $(N_F-1)$-tuple filter combinations.

\subsubsection{Step 3}
If $N_F = 1$, the optimum filter quality factor ($q^{opt}$) for each filter location is determined such that the highest reduction in $\operatorname{E}[{{S}_{\textit{THD}}}]$ is achieved. For multiple filter combinations, each filter quality factor is selected equal to the optimum value when the filter is singly used.

It should be noted that in order to maximize each individual filter efficiency in harmonic mitigation in case other filters are taken out of service, it is required to choose an optimum quality factor in single-filter scenarios. Additionally, it has been verified through comprehensive examination that the optimum quality factor for a filter in a specific location is generally the same under various filter placement scenarios.

\subsubsection{Step 4}
Since each harmonic filter injects reactive power to the grid, the network fundamental power flow pattern is modified. Here, the power flow constraints (bus voltage and branch current limits) are checked for each $N_F$-tuple filter combination in $A^{N_F}$. Initially, each filter capacity is assumed to be $Q = Q_{max}$. If the power flow constraints are violated, for all capacities combinations which satisfy the constraints, the system-wide index is calculated and an optimum combination is accordingly selected.

\subsubsection{Step 5}
Any $N_F$-tuple filter combination in $A^{N_F}$ causing a \textit{desirable} reduction in $\operatorname{E}[S_{\textit{THD}}]$ is stored in $B^{N_F}$, which can be mathematically expressed as
{\setlength{\mathindent}{0cm}\begin{equation}\label{eq.B}
\begin{gathered}
B^{N_F}=\{b^{N_F}\mid b^{N_F} \in A^{N_F}, \operatorname{E}[S_{\textit{THD}}(b^{N_F})] < p^{N_F-1}\}
\end{gathered}
\end{equation}}

\noindent where $p^{N_F-1}$ can theoretically be infinite, however a finite value is selected in order to restrict the search space to effective filter combinations. $p^0$ is selected as the first $d^0$-quantile of $A^1$, and for $n\geq 1$, $p^{n}$ is defined as the first $d^{n}$-quantile of $A^{n}$.

$d^n$ should be wisely selected by the operator to limit the computation burden while finding the optimum solution. Since $B^{N_F}$ is the desirable subset of $A^{N_F}$, which itself is built using $B^{N_F-1}$, not only any member of $B^{N_F}$ corresponds to a desirable $N_F$-tuple filter combination, but also any $n$-tuple filter combination subset of $b^{N_F}$ is desirable.

The members of $B^{N_F}$ are sorted based on $\operatorname{E}[{{S}_{\textit{THD}}}(b^{{{N}_{F}}})]$ in an ascending order.

\subsubsection{Step 6}
The members of $B^{N_F}$ are analyzed one by one to find the filter combination that satisfies the harmonic standard limits. Once such a filter combination is found, the algorithm is terminated. If no such filter combination is found, $N_F$ is incremented and the algorithm returns to \textbf{Step 2}.

\section{Simulation and Results}
\label{S5}

\subsection{Test System}
The strategy is implemented in MATLAB and applied to IEEE 118-bus transmission system \cite{washington}. The system has two voltage levels of 138 kV and 345 kV. Only 138 kV buses are candidate filter locations due to economic reasons and filter design complexities at 345 kV. It is assumed that the maximum reactive capacity for each harmonic filter is 20 MVAR, and the fundamental voltage in none of the buses should exceed 1.05 pu. Each transformer is considered to have the least rating multiple of 50 MVA which is larger than the apparent power of the connected load with an equivalent reactance of $X_{\textit{TR}} = 0.13\textnormal{ }pu$. The ratio of the apparent power of nonlinear loads to the total load is deemed $K_E = 0.2$.

The current harmonic distortion of each nonlinear load in the system is such that the 95\textsuperscript{th} percentile of the total demand distortion (TDD) matches the standard limits expressed in \hyperref[table.IHD_limits]{Table~\ref*{table.IHD_limits}}. Considering the beta distribution for the current IHDs, the means and standard deviations expressed in \hyperref[table.Harmonics]{Table~\ref*{table.Harmonics}} are obtained for harmonic loads in various $I_{SC}/I_L$ ranges. The harmonic phase angles are assumed to be uniformly distributed between $0$ and $2\pi$. It should be noted that zero sequence load current cannot find its way through the wye-delta transformer to the HV system. It is inferred that triplen (i.e. 3\textsuperscript{rd}, 9\textsuperscript{th}, etc.) harmonic currents available at the HV level are in fact positive/negative sequence triplens which are generated due to unsymmetrical loading conditions. As a result, only positive/negative sequence impedances are of interest for harmonic studies at HV.

\begin{table}
	\renewcommand{\arraystretch}{1.3}
	\caption{Current distortion limits for $69kV<V_{rated}<161kV$ \cite{IEEE519}}
	\label{table.IHD_limits}
	\centering
	\begin{tabular}{|c|c|c|}
		\hline
		\textbf{$I_{SC}/I_L$} & $3 \leq h < 11$ & TDD \\
		\hline
		$< 20$ & 2.0 & 2.5 \\
		\hline
		$20<50$ & 3.5 & 4.0 \\
		\hline
		$50<100$ & 5.0 & 6.0 \\
		\hline
		$100<1000$ & 6.0 & 7.5 \\
		\hline
		$>1000$ & 7.5 & 10.0 \\
		\hline
	\end{tabular}
\end{table}

\begin{table}
	\renewcommand{\arraystretch}{1.3}
	\setlength\tabcolsep{1.5pt}
	\caption{Nonlinear load harmonics means and standard deviations in percentage of fundamental current}
	\label{table.Harmonics}
	\centering
	\begin{tabular}{|c|c|c|c|c|c|c|}
		\hline
		$I_{SC}/I_L$&	$\operatorname{E}[\alpha^{(3)}]$&	$\operatorname{SD}[\alpha^{(3)}]$&	$\operatorname{E}[\alpha^{(5)}]$&	$\operatorname{SD}[\alpha^{(5)}]$&	$\operatorname{E}[\alpha^{(7)}]$&	$\operatorname{SD}[\alpha^{(7)}]$\\
		\hline
		\textless20&	0.20\%&	0.11\%&	1.00\%&	0.53\%&	0.72\%&	0.38\%\\
		\hline
		20\textless50&	0.32\%&	0.17\%&	1.61\%&	0.85\%&	1.15\%&	0.61\%\\
		\hline
		50\textless100&	0.48\%&	0.26\%&	2.41\%&	1.28\%&	1.72\%&	0.91\%\\
		\hline
		100\textless1000&	0.60\%&	0.32\%&	3.01\%&	1.60\%&	2.15\%&	1.14\%\\
		\hline
	\end{tabular}
\end{table}

\subsection{Base Case}
The algorithm is executed for the base case (Case 0), and the buses introduced in \hyperref[table.V95init]{Table~\ref*{table.V95init}} are found to violate the standard limits.

\begin{table}
	\renewcommand{\arraystretch}{1.55}
	\setlength\tabcolsep{2.5pt}
	\caption{$\tilde{V}_{\textit{HD},95\%}$ of buses violating the standard limits in Case 0}
	\label{table.V95init}
	\centering
	\begin{tabular}{|c|c|c|c|}
		\hline
		Bus	&$\tilde{V}_{\textit{IHD},95\%}^{(5)}$&	$\tilde{V}_{\textit{IHD},95\%}^{(7)}$&	$\tilde{V}_{\textit{THD},95\%}$ \\ \hline
		9	&1.14\%	&0.24\%	&1.15\% \\ \hline
		43	&1.68\%	&1.15\%	&1.84\% \\ \hline
		44	&2.52\%	&1.97\%	&2.89\% \\ \hline
		45	&2.11\%	&1.21\%	&2.24\% \\ \hline
		82	&1.14\%	&1.85\%	&1.98\% \\ \hline
		83	&1.13\%	&1.94\%	&2.07\% \\ \hline
	\end{tabular}
\end{table}

For validation, MCS is performed with one million samples. Considering gamma and log-normal distributions for $V_{\textit{THD},i}^2$, the MLF parameters are calculated. The corresponding log-likelihoods of MLF and EF for buses 79 and 82 are reported in \hyperref[table.L_V]{Table~\ref*{table.L_V}}. Comparing the log-likelihoods shows that the gamma distribution fits the data better, and the likelihood of the EF is very close to the maximum likelihood. It is notable that for all harmonic voltages of buses, gamma distribution yields higher values of 95\textsuperscript{th} percentiles which also better matches the MCS data.

\begin{table}
	\renewcommand{\arraystretch}{1.3}
	\caption{Log-likelihoods of various fits to $V_{\textit{THD},i}^2$ (Case 0)}
	\label{table.L_V}
	\centering
	\begin{tabular}{|c|c|c|c|}
		\hline
		$i$& PDF &	MLF&	EF\\
		\hline
		\multirow{2}{*}{79}&Gamma &	$8.651\times 10^{6}$&	$8.651\times 10^{6}$\\
		& Log-normal &	$8.613\times 10^{6}$&	$8.555\times 10^{6}$\\
		\hline
		\multirow{2}{*}{82}&Gamma &	$7.847\times 10^{6}$&	$7.845\times 10^{6}$\\
		& Log-normal&	$7.819\times 10^{6}$&	$7.767\times 10^{6}$\\
		\hline
	\end{tabular}
\end{table}

The properties mean, variance, and the 95\textsuperscript{th} percentile for the MCS data, the gamma EF, and the gamma MLF are calculated. These properties for buses 79 and 82 are expressed in \hyperref[table.V]{Table~\ref*{table.V}}. It can be seen that the properties of the EF accurately match those of MCS.

\begin{table}
	\renewcommand{\arraystretch}{1.3}
	\caption{Properties of $V_{\textit{THD},i}^2$ calculated using different methods (Case 0)}
	\label{table.V}
	\centering
	\setlength\tabcolsep{2.5pt}
	\begin{tabular}{|c|c|c|c|c|}
		\hline
		$i$ & Method &	Mean&	Variance&	95\textsuperscript{th} Percentile\\
		\hline
		\multirow{3}{*}{79}&MCS&	$7.189\times 10^{-5}$&	$2.713\times 10^{-9}$&	$1.728\times 10^{-4}$\\
		&EF&	$7.186\times 10^{-5}$&	$2.715\times 10^{-9}$&	$1.732\times 10^{-4}$\\
		&MLF&	$7.189\times 10^{-5}$&	$2.632\times 10^{-9}$&	$1.715\times 10^{-4}$\\
		\hline
		\multirow{3}{*}{82}&MCS&	$15.604\times 10^{-5}$&	$15.022\times 10^{-9}$&	$3.949\times 10^{-4}$\\
		&EF&	$15.606\times 10^{-5}$&	$14.994\times 10^{-9}$&	$3.959\times 10^{-4}$\\
		&MLF&	$15.604\times 10^{-5}$&	$13.795\times 10^{-9}$&	$3.852\times 10^{-4}$\\
		\hline
	\end{tabular}
\end{table}

\subsection{Filter Placement}
\begin{figure}
	\centering
	\includegraphics[width=3.2in]{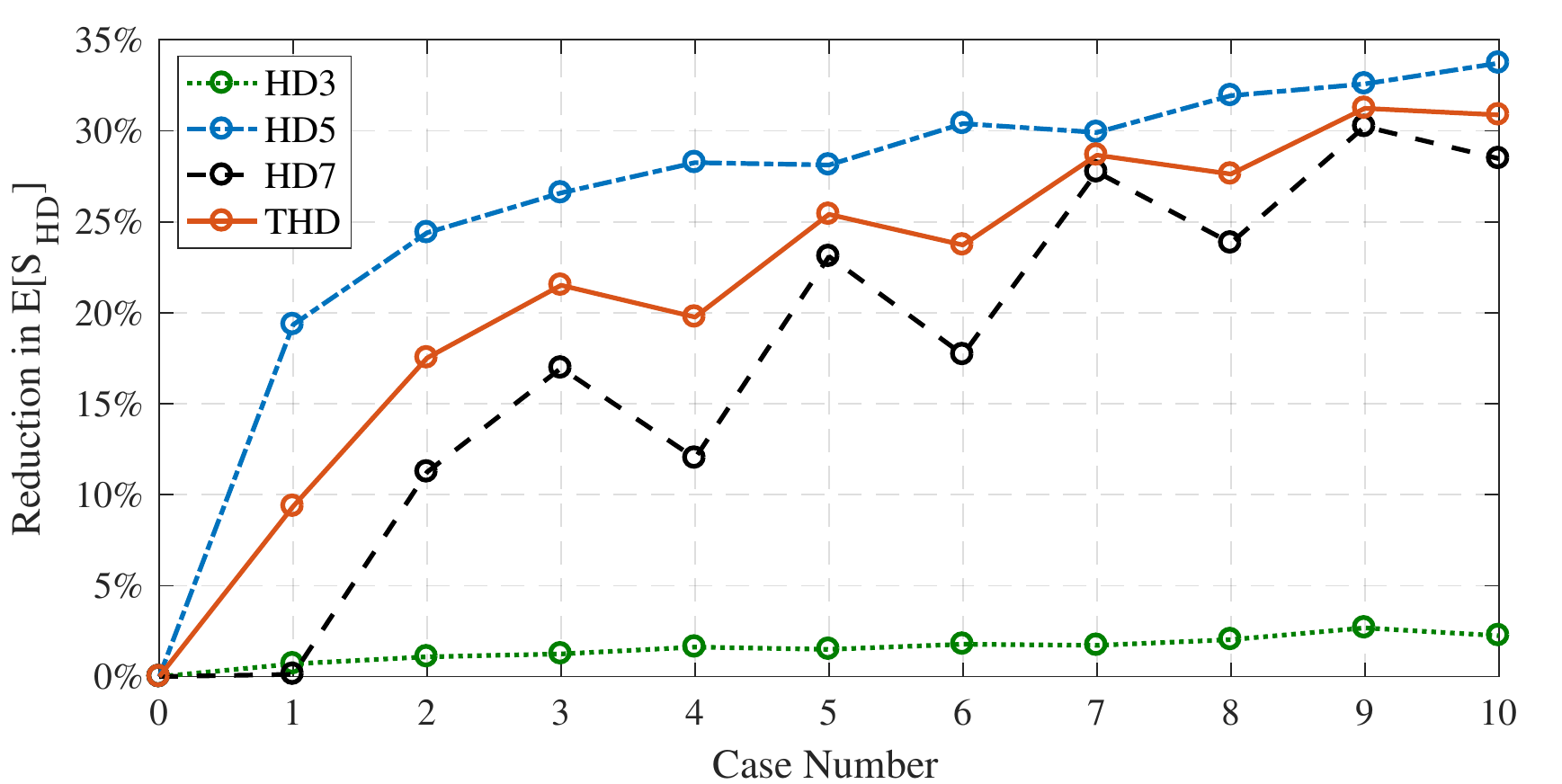}
	\caption{Reduction in $\operatorname{E}[S_{\textit{HD}}]$ for different cases.}
	\label{fig.S_NF}
\end{figure}

As some harmonic voltages exceed the standard threshold in the base case, the algorithm proceeds to find acceptable filter combinations. In the first step, the optimum quality factor for all filter locations are determined and their values are shown in \hyperref[table.q]{Table~\ref*{table.q}} for several buses. Only for $N_{F}\geq 3$ there are combinations that meet the standards. In \hyperref[table.Cases]{Table~\ref*{table.Cases}} and \hyperref[fig.S_NF]{Fig.~\ref*{fig.S_NF}}. The results of several cases with $1\leq N_F \leq 6$ are illustrated.  Among the 3-tuple filter combinations, Case 3 is the one with minimum $\operatorname{E}[S_{\textit{THD}}]$, whereas Case 4 is the best one that satisfies the standard. Case 4 is considered as the final solution. However, the results for further numbers of filters are studied for comparison. It is noteworthy that since the fundamental voltage limit is not exceeded, the capacity for all filters is 20 MVAR.

\begin{table}
	\renewcommand{\arraystretch}{1.3}
	\caption{Optimum quality factor for several filter locations}
	\label{table.q}
	\centering
	\begin{tabular}{|c|c|c|c|}
		\hline
		Bus & $q^{opt}$ & Bus &  $q^{opt}$ \\
		\hline
		21 & 1.3 & 48 & 1.0 \\
		\hline
		34 & 2.3 & 79 & 1.0 \\
		\hline
		44 & 1.0 & 82 & 1.0 \\
		\hline
		45 & 1.0 & 105 & 2.3 \\
		\hline
	\end{tabular}
\end{table}

\begin{table}
	\renewcommand{\arraystretch}{1.3}
	\caption{Filter placement cases}
	\label{table.Cases}
	\centering
	\setlength\tabcolsep{2.5pt}
	\begin{tabular}{|c|c|c|c|c|}
		\hline
		Case&	$N_F$&	Filter Locations&	$\operatorname{E}[S_{\textit{THD}}]$&	\begin{tabular}{@{}c@{}}Standard\\ Satisfied?\end{tabular}\\
		\hline
		Case 0&	0&	$-$&	$3.88\times 10^{-5}$&	 \ding{55}\\
		\hline
		Case 1&	1&	$44$&	$3.52\times 10^{-5}$&	\ding{55}\\
		\hline
		Case 2&	2&	$44, 82$&	$3.20\times 10^{-5}$&	\ding{55}\\
		\hline
		Case 3&	3&	$44, 48, 82$&	$3.05\times 10^{-5}$&	\ding{55}\\
		\hline
		Case 4&	3&	$34, 44, 82$&	$3.11\times 10^{-5}$&	\ding{51}\\
		\hline
		Case 5&	4&	$44, 48, 82, 105$&	$2.89\times 10^{-5}$&	\ding{55}\\
		\hline
		Case 6&	4&	$34, 44, 48, 82$&	$2.96\times 10^{-5}$&	\ding{51}\\
		\hline
		Case 7&	5&	$44, 48, 79, 82, 105$&	$2.77\times 10^{-5}$&	\ding{55}\\
		\hline
		Case 8&	5&	$34, 44, 48, 82, 105$&	$2.81\times 10^{-5}$&	\ding{51}\\
		\hline
		Case 9&	6&	$44, 45, 48, 79, 82, 105$&	$2.67\times 10^{-5}$&	\ding{55}\\
		\hline
		Case 10&	6&	$34, 44, 48, 79, 82, 105$&	$2.68\times 10^{-5}$&	\ding{51}\\
		\hline
	\end{tabular}
\end{table}

\begin{figure}
	\centering
	\includegraphics[width=3.2in]{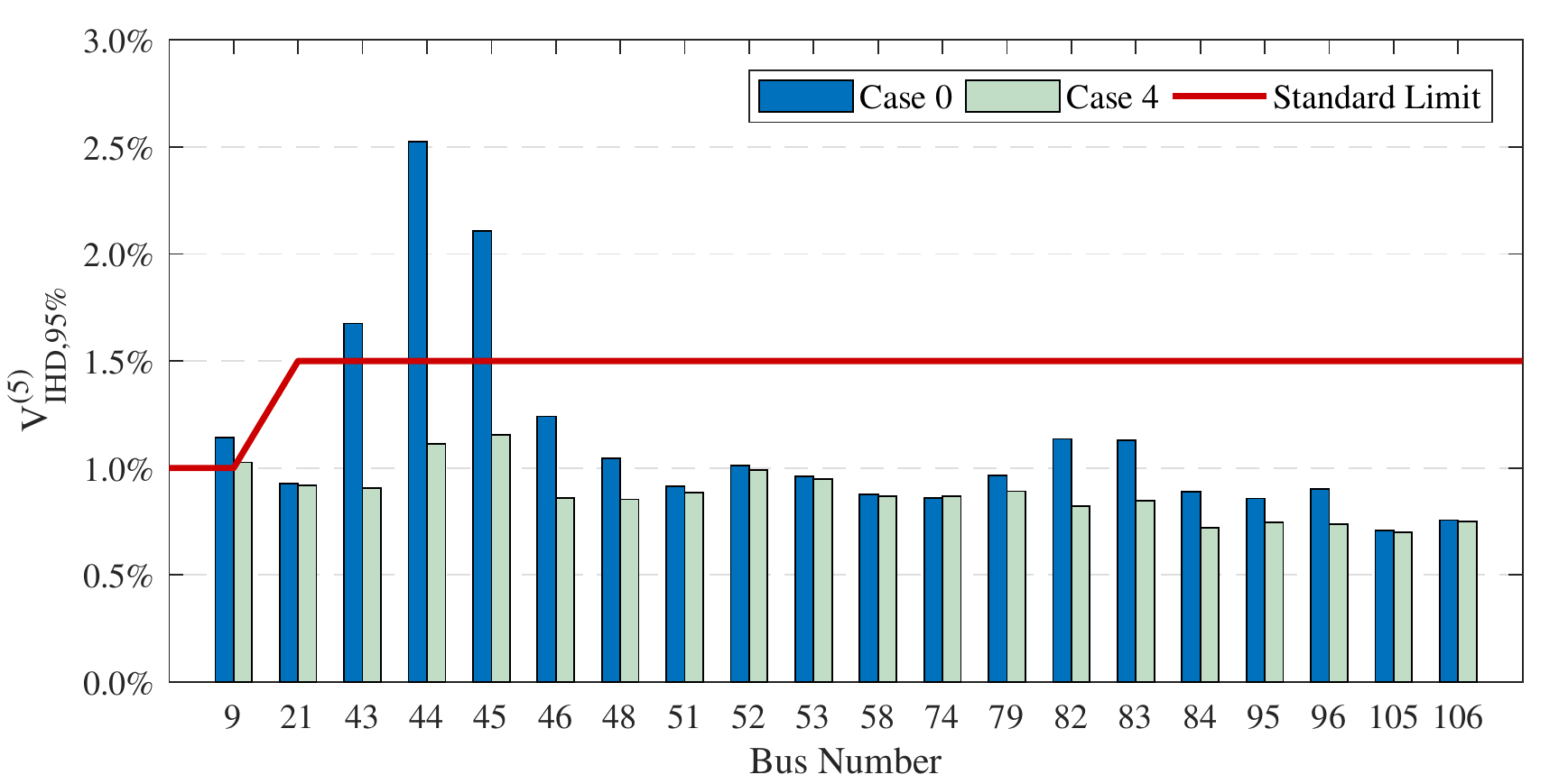}
	\caption{Voltage 5\textsuperscript{th} harmonic distortions at the buses before and after filter placement.}
	\label{fig.Vh5_C95}
\end{figure}

\begin{figure}
	\centering
	\includegraphics[width=3.2in]{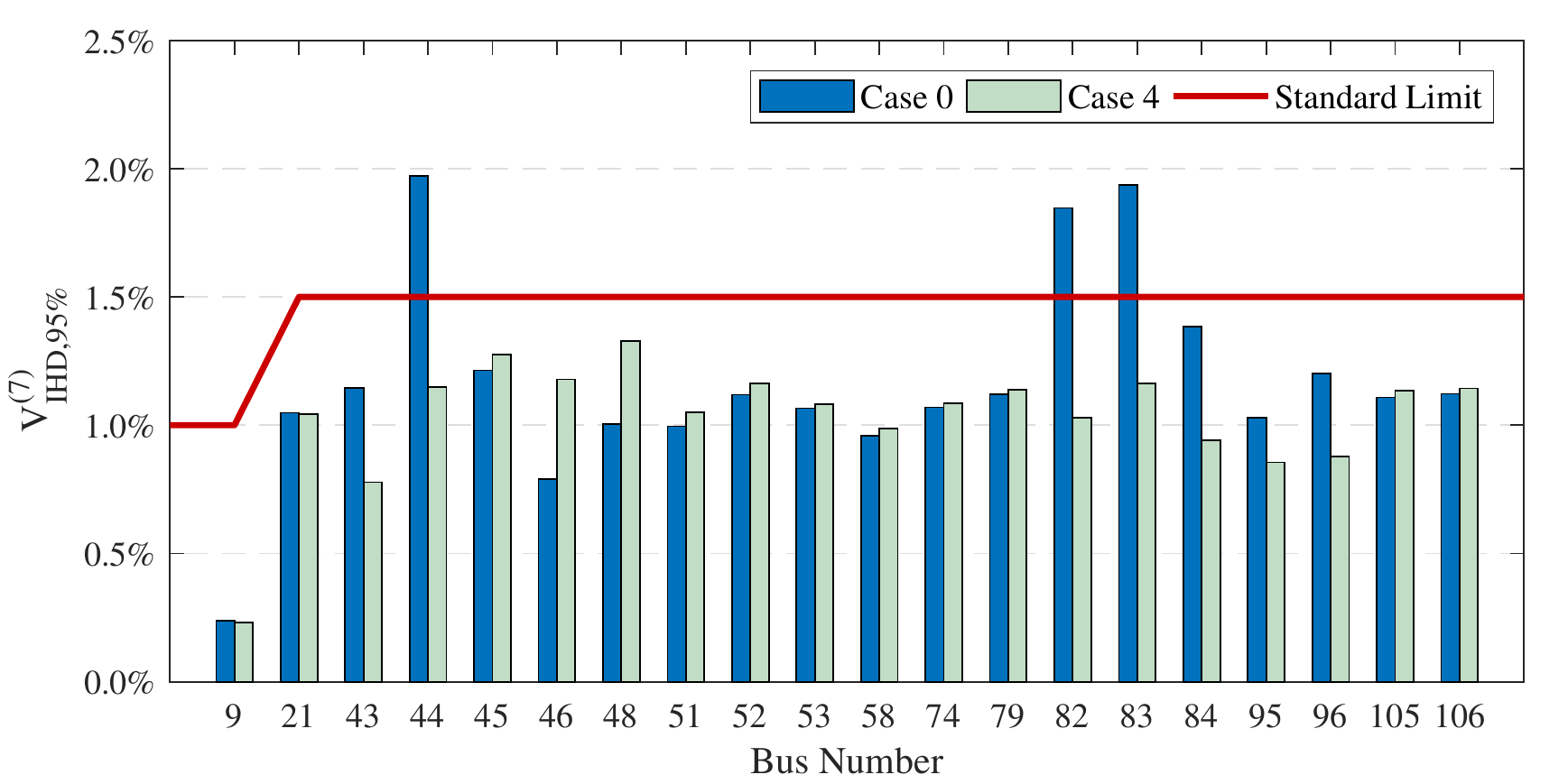}
	\caption{Voltage 7\textsuperscript{th} harmonic distortions at the buses before and after filter placement.}
	\label{fig.Vh7_C95}
\end{figure}

\begin{figure}
	\centering
	\includegraphics[width=3.2in]{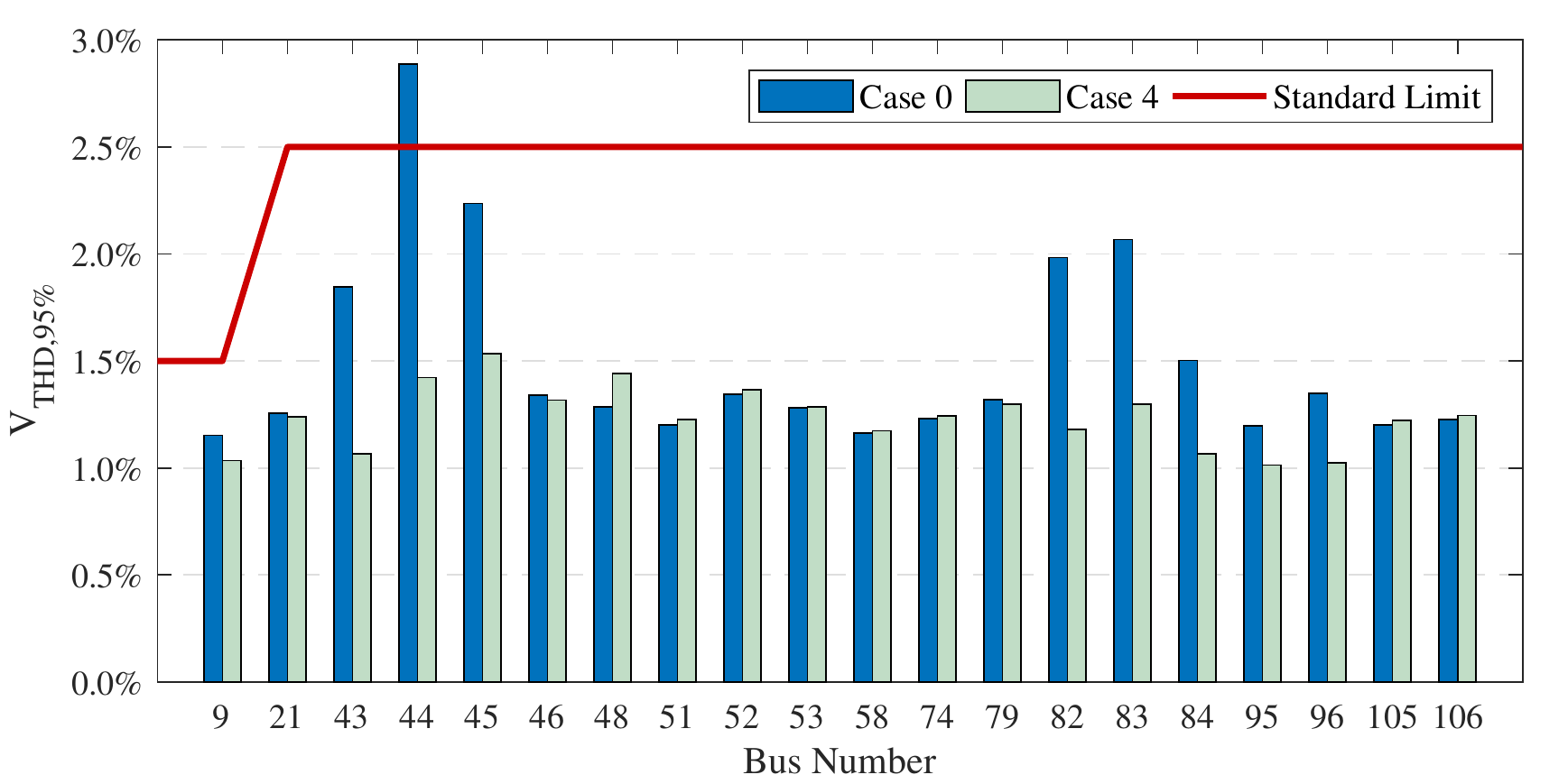}
	\caption{Voltage total harmonic distortions at the buses before and after filter placement.}
	\label{fig.Vthd_C95}
\end{figure}

Figs.~\ref{fig.Vh5_C95} - \ref{fig.Vthd_C95} depict the estimated 95\textsuperscript{th} percentiles of voltage distortions in Cases 0 and 4 for 20 buses with the maximum initial THD values.

To verify the validity of the results, MCS is executed for Case 4. \hyperref[table.L_Vp]{Table~\ref*{table.L_Vp}} shows that the gamma PDF is superior in terms of likelihood and \hyperref[table.Vp]{Table~\ref*{table.Vp}} indicates that the EF approximates the data set as precisely as the MLF.

\begin{table}
	\renewcommand{\arraystretch}{1.3}
	\caption{Log-likelihoods of various fits to $V_{\textit{THD},i}^2$ (Case 4)}
	\label{table.L_Vp}
	\centering
	\begin{tabular}{|c|c|c|c|}
		\hline
		$i$& PDF &	MLF&	EF\\
		\hline
		\multirow{2}{*}{79}&Gamma&	$8.687\times 10^{6}$&	$8.686\times 10^{6}$\\
		&Log-normal&	$8.653\times 10^{6}$&	$8.599\times 10^{6}$\\
		\hline
		\multirow{2}{*}{82}&Gamma&	$8.871\times 10^{6}$&	$8.871\times 10^{6}$\\
		&Log-normal&	$8.836\times 10^{6}$&	$8.779\times 10^{6}$\\
		\hline
	\end{tabular}
\end{table}

\begin{table}
	\renewcommand{\arraystretch}{1.3}
	\caption{Properties of $V_{\textit{THD},i}^2$ calculated using different methods (Case 4)}
	\label{table.Vp}
	\centering
	\setlength\tabcolsep{2.5pt}
	\begin{tabular}{|c|c|c|c|c|}
		\hline
		$i$ & Method &	Mean &	Variance&	95\textsuperscript{th} Percentile\\
		\hline
		\multirow{3}{*}{79}&MCS&	$6.889\times 10^{-5}$&	$2.598\times 10^{-9}$&	$1.677\times 10^{-4}$\\
		&EF&	$6.886\times 10^{-5}$&	$2.600\times 10^{-9}$&	$1.682\times 10^{-4}$\\
		&MLF&	$6.889\times 10^{-5}$&	$2.473\times 10^{-9}$&	$1.656\times 10^{-4}$\\
		\hline
		\multirow{3}{*}{82}&MCS&	$5.752\times 10^{-5}$&	$1.770\times 10^{-9}$&	$1.389\times 10^{-4}$\\
		&EF&	$5.753\times 10^{-5}$&	$1.768\times 10^{-9}$&	$1.394\times 10^{-4}$\\
		&MLF&	$5.752\times 10^{-5}$&	$1.702\times 10^{-9}$&	$1.377\times 10^{-4}$\\
		\hline
	\end{tabular}
\end{table}

\begin{figure}
	\centering
	\includegraphics[width=3.2in]{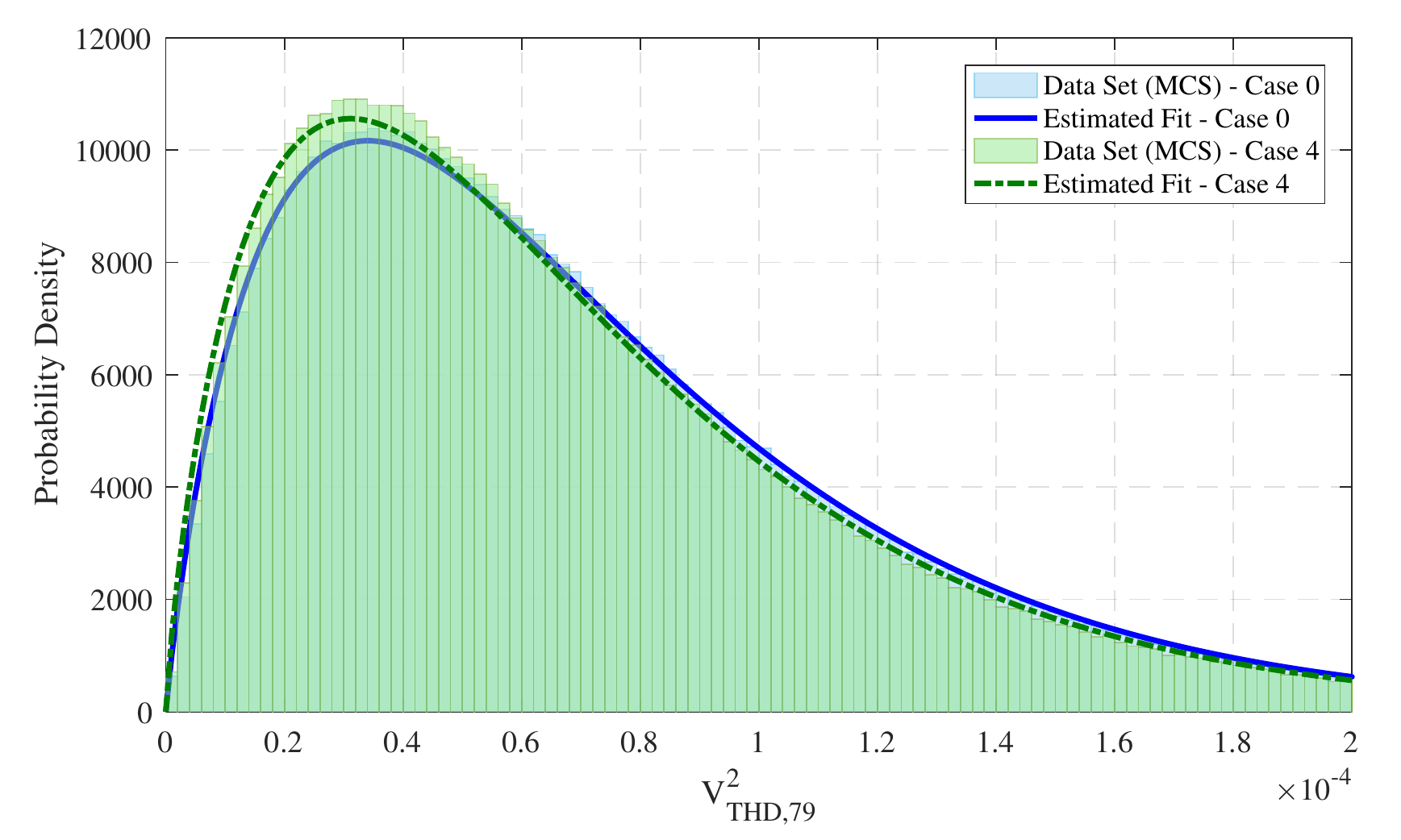}
	\caption{Probability density of $V_{\textit{THD},79}^2$.}
	\label{fig.V2thd79}
\end{figure}

\begin{figure}
	\centering
	\includegraphics[width=3.2in]{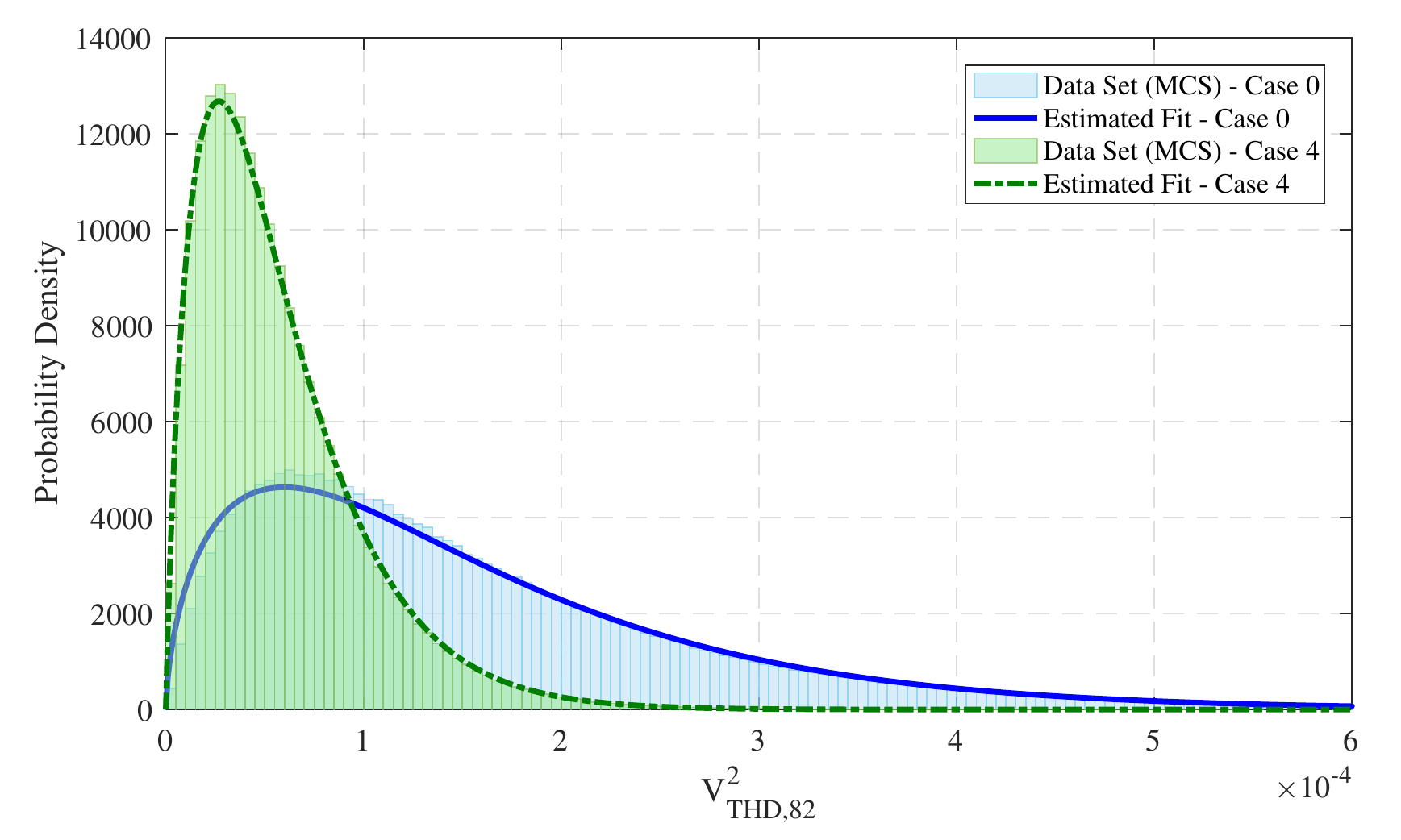}
	\caption{Probability density of $V_{\textit{THD},82}^2$.}
	\label{fig.V2thd82}
\end{figure}

Figs.~\ref{fig.V2thd79} and~\ref{fig.V2thd82} show the probability density of the squared voltage total harmonic distortion of buses 79 and 82 respectively in Cases 0 and 4. It is concluded that the EF predicts the trend of PDF accurately.

\begin{figure}
	\centering
	\includegraphics[width=3.2in]{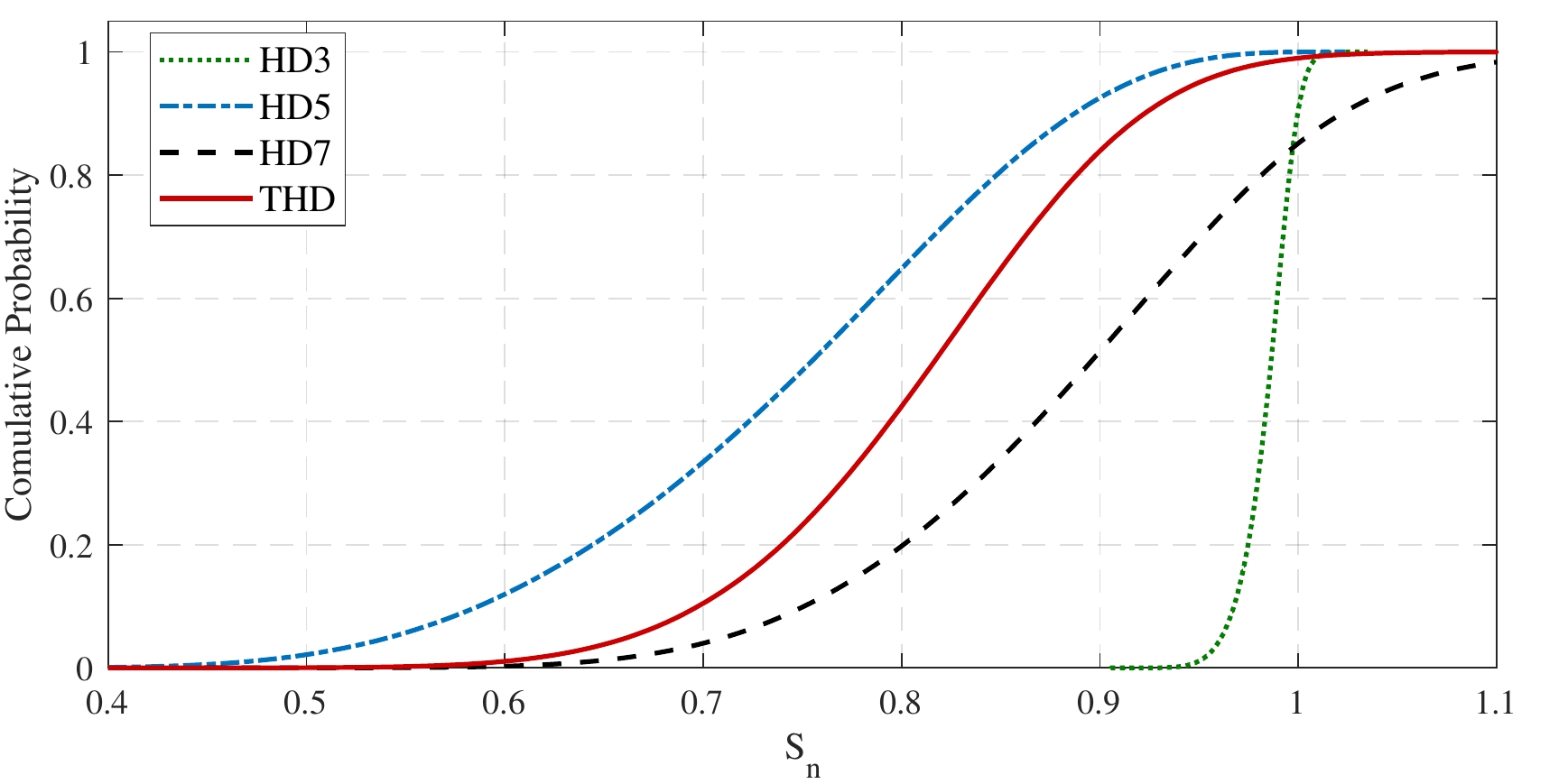}
	\caption{Cumulative density of $\frac{S_{\textit{HD}}(\textnormal{Case 4})}{S_{\textit{HD}}(\textnormal{Case 0})}$.}
	\label{fig.Sn}
\end{figure}

To evaluate the uncertainty involved in the filter placement, the cumulative density functions of $\frac{S_{\textit{HD}}(\textnormal{Case 4})}{S_{\textit{HD}}(\textnormal{Case 0})}$ for individual and total harmonic distortions are extracted using MCS and illustrated in \hyperref[fig.Sn]{Fig.~\ref*{fig.Sn}}. Defining the risk as the probability that the system-wide index increases with regard to the base case, the risk for Case 4 at the 3\textsuperscript{rd}, 5\textsuperscript{th}, 7\textsuperscript{th}, and total harmonics are 9.38\%, 0.01\%, 14.82\%, and 1.00\% respectively.

\subsection{Harmonic Modal Analysis}
\begin{figure}
	\centering
	\includegraphics[width=3.2in]{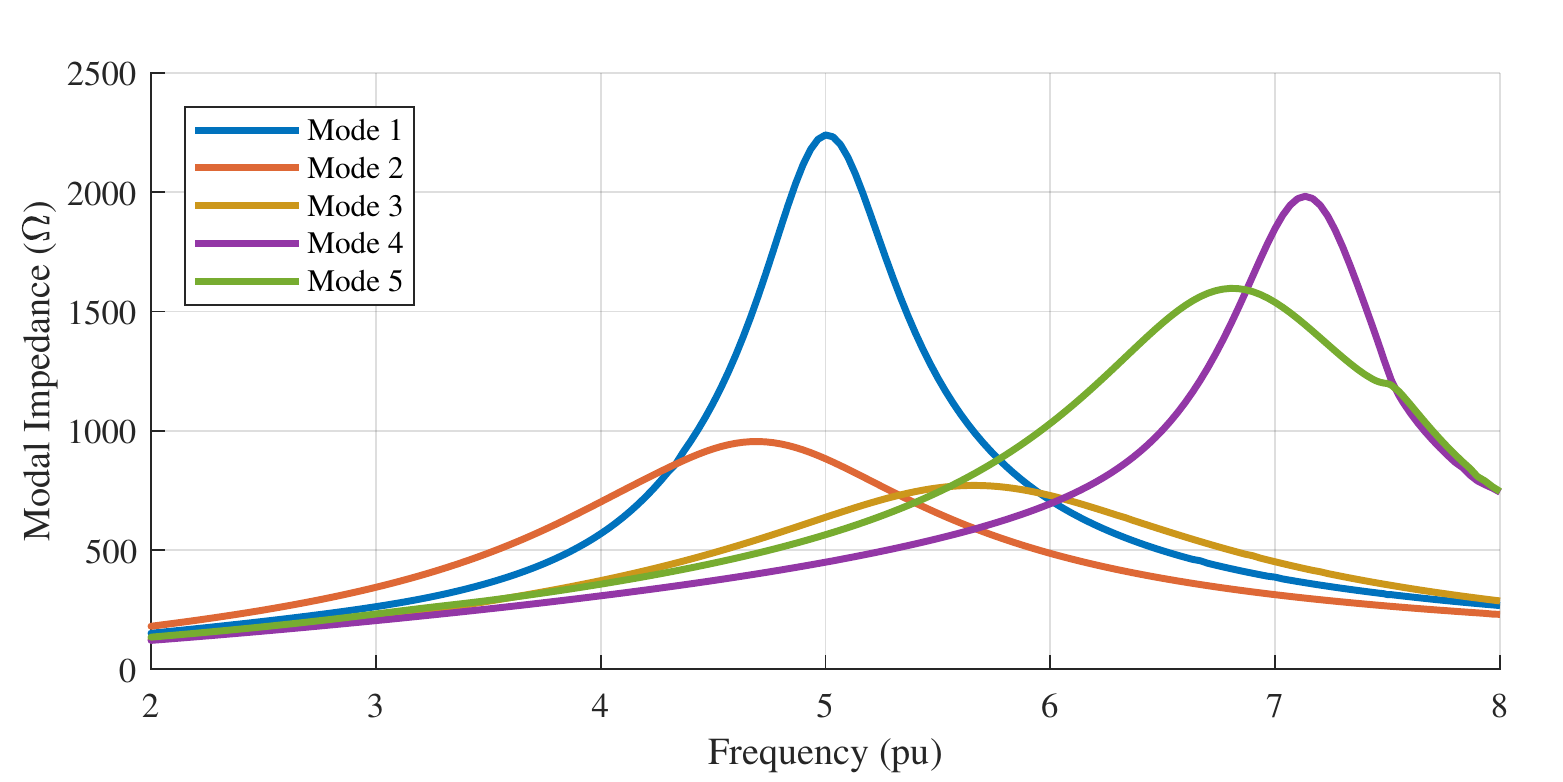}
	\caption{Critical modal impedances at harmonics 5 and 7 (Case 0).}
	\label{fig.ZCase0}
\end{figure}

\begin{figure}
	\centering
	\includegraphics[width=3.2in]{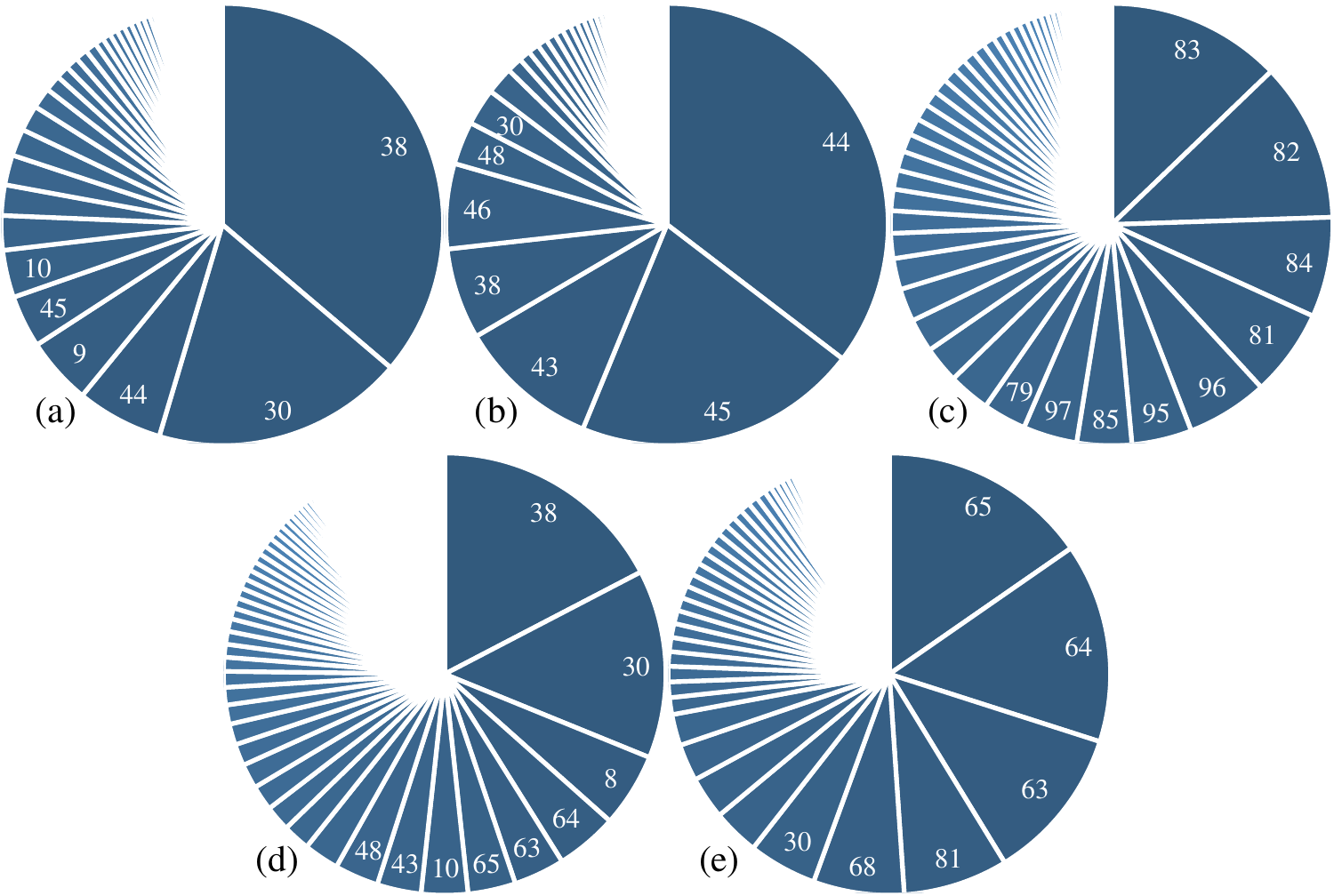}
	\caption{Bus participation factors in critical modes: (a) mode 1 at 5\textsuperscript{th} harmonic, (b) mode 2 at 5\textsuperscript{th} harmonic, (c) mode 3 at 5\textsuperscript{th} harmonic, (d) mode 4 at 7\textsuperscript{th} harmonic, and (e) mode 5 at 7\textsuperscript{th} harmonic.}
	\label{fig.PFs}
\end{figure}

To further investigate the effectiveness of the filter placement strategy, the critical modes are found with the method explained in \cite{Xu2005}. The corresponding modal impedances are plotted in \hyperref[fig.ZCase0]{Fig.~\ref*{fig.ZCase0}} for a frequency range of 120-480 Hz. Bus participation factors in these modes are shown in \hyperref[fig.PFs]{Fig.~\ref*{fig.PFs}}.

It is observed that critical modes 1, 4, and 5 are primarily correlated with 345 kV buses, such as buses 30, 38, 64, and 65, which are not filter placement candidates. Conversely, modes 2 and 3 are chiefly correlated with 138 kV buses, such as buses 44, 45, 82, and 83, among which buses 44 and 82 are previously found as effective filter locations. However buses 34, 48, 79, and 105 which are promising filter locations are not suggested by the modal method partly because the modal analysis ignores the current injection magnitudes.

\begin{figure}
	\centering
	\includegraphics[width=3.2in]{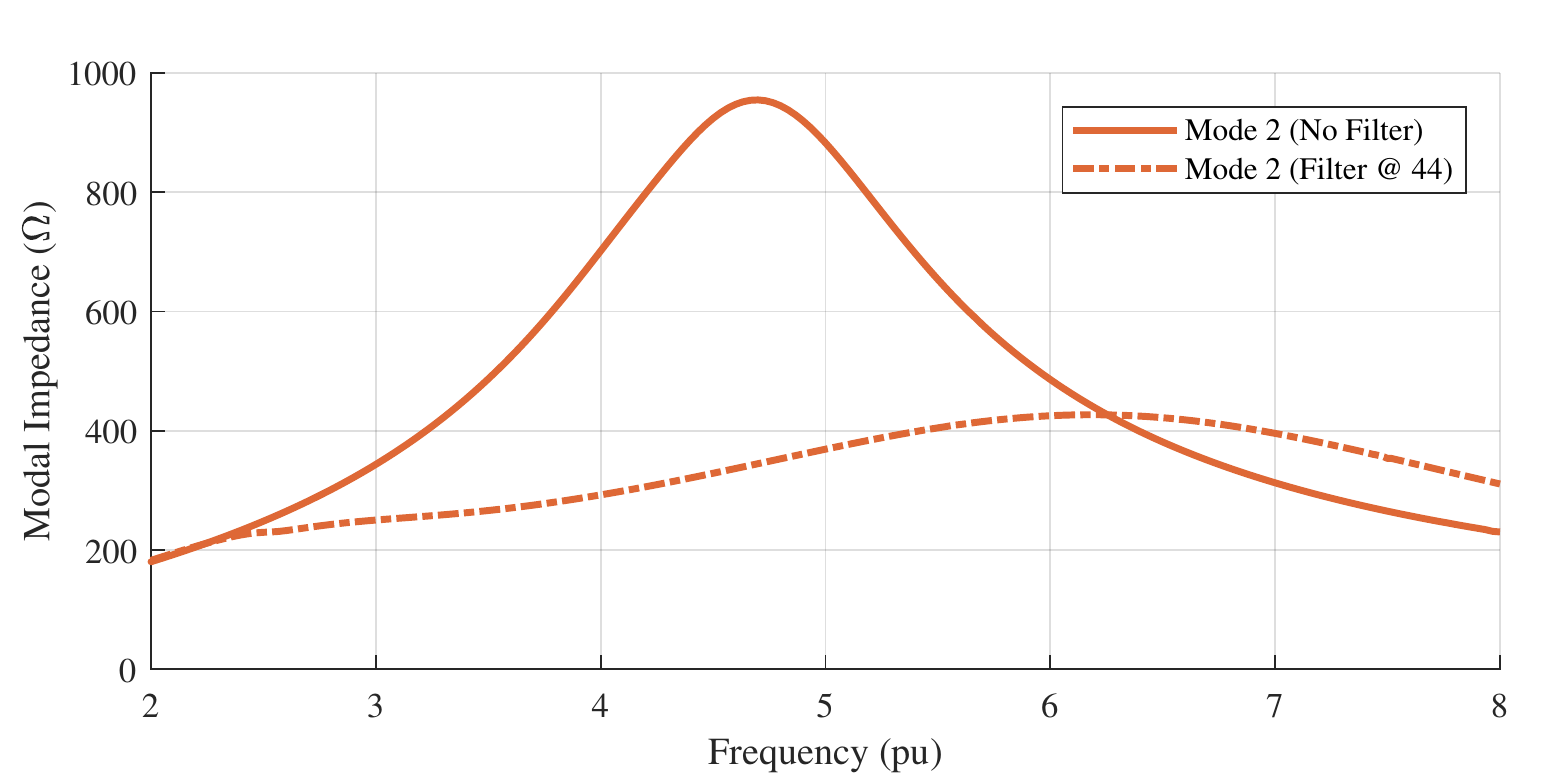}
	\caption{Effect of filter placement at bus 44 on critical mode 2.}
	\label{fig.ZF44}
\end{figure}

\begin{figure}
	\centering
	\includegraphics[width=3.2in]{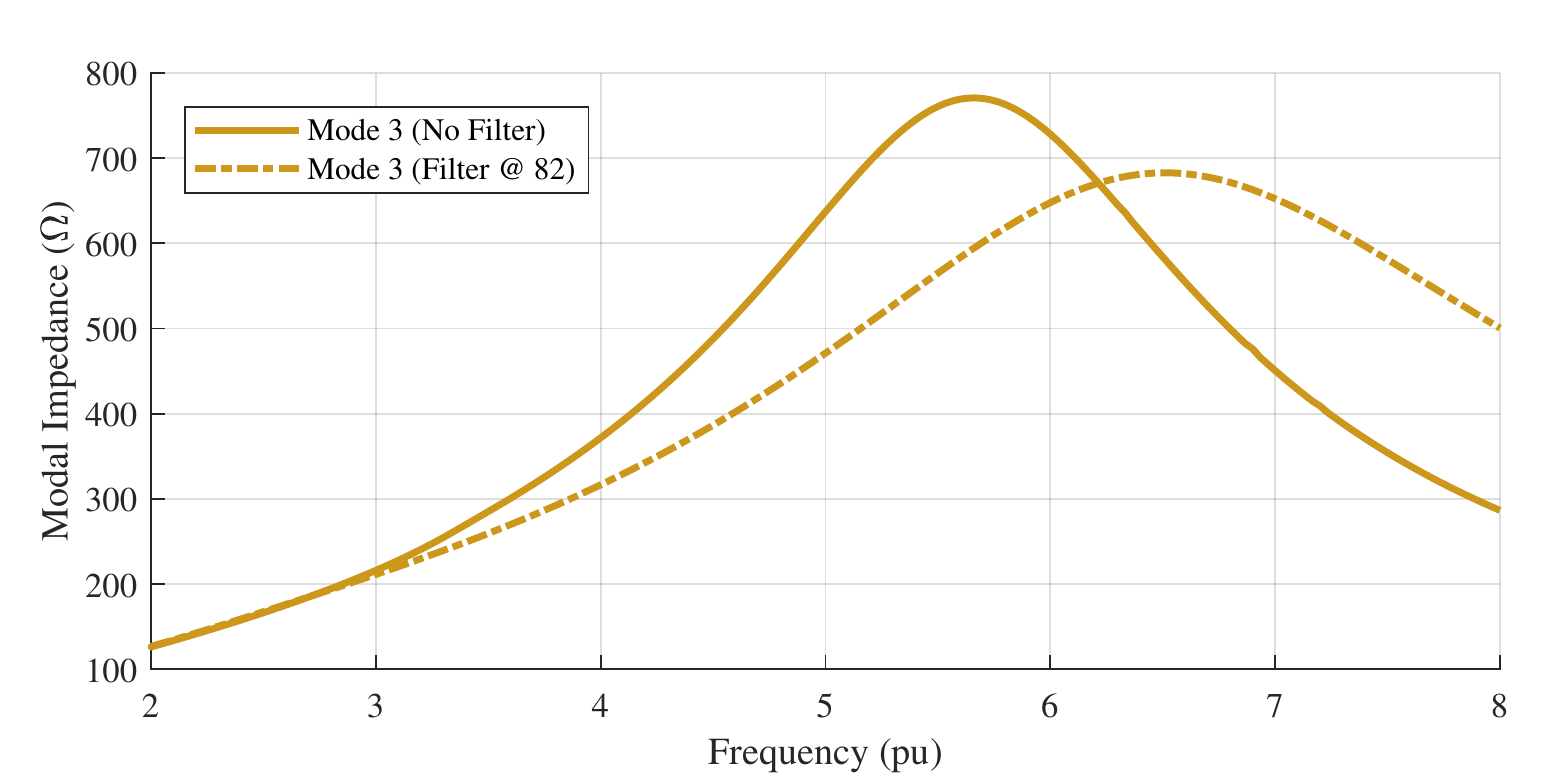}
	\caption{Effect of filter placement at bus 82 on critical modes 3 and 5.}
	\label{fig.ZF82}
\end{figure}

The effect of installing harmonic filters at buses 44 and 82 on the critical modes is illustrated in \hyperref[fig.ZF44]{Fig.~\ref*{fig.ZF44}} and \hyperref[fig.ZF82]{Fig.~\ref*{fig.ZF82}} when each filter is installed individually. Analysis of   \hyperref[fig.ZF44]{Fig.~\ref*{fig.ZF44}} and \hyperref[fig.ZF82]{Fig.~\ref*{fig.ZF82}} reveals that the selected filter combination can alleviate menacing harmonic resonances in the test transmission system.

\section{Conclusion}
\label{S6}
A passive filter placement strategy in transmission systems in the presence of stochastically modeled loads is presented in this paper. The analytical strategy can be employed in planning procedures when no measurements are available and in wide-area systems with sparse information about harmonic distortions of loads. The strategy seeks to minimize the expected value of a system-wide harmonic index, while respecting the fundamental power flow constraints and the voltage harmonic standard limits. The hierarchical filter placement algorithm employs a computationally efficient method, that both restricts the search space to effective filter combinations and ensures any subset of the final filter solution is influential in harmonic mitigation. Harmonic modal analysis confirms that this approach is capable of relieving resonances in the system frequency response.

\footnotesize
\bibliographystyle{IEEEtran}
\bibliography{IEEEabrv,References}

\end{document}